\documentclass[fleqn,usenatbib]{mnras}

\usepackage[T1]{fontenc}

\DeclareRobustCommand{\VAN}[3]{#2}
\let\VANthebibliography\thebibliography
\def\thebibliography{\DeclareRobustCommand{\VAN}[3]{##3}\VANthebibliography}

\usepackage{graphicx}	
\usepackage{amsmath}	
\usepackage{amssymb}	
\usepackage{xcolor}
\usepackage{lipsum}
\usepackage[english]{babel}
\usepackage{blindtext}
\usepackage{scalefnt}
\usepackage{soul}
\usepackage{multirow}
\usepackage{makecell}
\usepackage{arydshln}
\usepackage{float}
\usepackage{siunitx}
\usepackage{booktabs}
\usepackage{caption}

\newcommand{\Lbol}{$L_{\rm{AGN}}$}
\newcommand{\LAGN}{$L_{\rm{AGN}}$}
\newcommand{\vAGN}{$v_{\rm{AGN}}$}
\newcommand{\vmin}{$v_{\rm{min}}$}

\newcommand{\kmin}{$k_{\rm{min}}$}
\newcommand{\Lmax}{$\lambda_{\rm{max}}$}
\newcommand{\Mtarget}{$M_{\rm{target}}$}
\newcommand{\Ek}{\ensuremath{\dot{E}_k/L}}

\newcommand{\LbolEq}[2]{\ensuremath{L_{\rm{AGN}}{#1} 10^{#2}\rm{\ erg\ s^{-1}}}}
\newcommand{\LAGNEq}[2]{\ensuremath{L_{\rm{AGN}}{#1} 10^{#2}\rm{\ erg\ s^{-1}}}}

\newcommand{\kminEq}[2]{\ensuremath{k_{\rm{min}}{#1}{#2}\rm{\ kpc^{-1}}}}
\newcommand{\LmaxEq}[2]{\ensuremath{\lambda_{\rm{max}}{#1}{#2}{\ \rm{pc}}}}
\newcommand{\MtargetEq}[2]{\ensuremath{M_{\rm{target}}{#1}{#2}{\ \rm{M_\odot}}}}
\newcommand{\vAGNEq}[2]{\ensuremath{v_{\rm{AGN}}{#1}{#2}{\rm{\ km\ s^{-1}}}}}
\newcommand{\vrEq}[2]{\ensuremath{v_{\rm{r}}{#1}{#2}{\rm{\ km\ s^{-1}}}}}
\newcommand{\vminEq}[2]{\ensuremath{v_{\rm{min}}{#1}{#2}{\rm{\ km\ s^{-1}}}}}
\newcommand{\TEq}[2]{\ensuremath{T{#1}10^{#2}{\rm{\ K}}}}
\newcommand{\tEq}[2]{\ensuremath{t{#1}{#2}{\rm{\ Myr}}}}

\newcommand{\pEq}[2]{\ensuremath{\dot{p}/(L/c){#1}{#2}}}
\newcommand{\EEq}[2]{\ensuremath{\dot{E}_k/L{#1}{#2}}\%}
\newcommand{\MofEq}[2]{\ensuremath{\dot{M}_{\rm{OF}}{#1}{#2}{\rm{\ M_\odot\ yr^{-1}}}}}

\definecolor{green}{RGB}{34, 139, 204}

\usepackage{newtxtext,newtxmath}

\title[Multiphase Outflows from AGN Winds]{AGN-driven outflows in clumpy media: multiphase structure and scaling relations}

\author[S.R. Ward et al.]{
S. R. Ward,$^{1,2,3,4}$\thanks{E-mail: \url{samuelruthvenward@gmail.com}}
T. Costa,$^{1,5}$\thanks{E-mail: \url{tiago.costa@newcastle.ac.uk}}
C. M. Harrison$^{1}$ and 
V. Mainieri$^{2}$
\\
$^{1}$School of Mathematics, Statistics and Physics, Newcastle University, Newcastle upon Tyne, NE1 7RU, UK \\
$^{2}$European Southern Observatory, Karl-Schwarzschild-Straße 2, 85748 Garching bei München, Germany\\
$^{3}$Excellence Cluster ORIGINS, Boltzmannstraße 2, 85748 Garching bei München, Germany\\
$^{4}$Ludwig Maximilian Universität, Professor-Huber-Platz 2, 80539 München, Germany \\
$^{5}$Max-Planck Institut für Astrophysik, Karl-Schwarzschild-Straße 1, 85748 Garching bei München, Germany
\\
}

\date{Accepted XXX. Received YYY; in original form ZZZ}

\pubyear{2024}

\begin{document}
\label{firstpage}
\pagerange{\pageref{firstpage}--\pageref{lastpage}}
\maketitle

\begin{abstract}

Small-scale winds driven from accretion discs surrounding active galactic nuclei (AGN) are expected to launch kpc-scale outflows into their host galaxies. However, the ways in which the structure of the interstellar medium (ISM) affects the multiphase content and impact of the outflow remains uncertain. We present a series of numerical experiments featuring a realistic small-scale AGN wind with velocity $5\times 10^3 \-- 10^4\rm{\ km\ s^{-1}}$ interacting with an isolated galaxy disc with a manually-controlled clumpy ISM, followed at sub-pc resolution. Our simulations are performed with \textsc{Arepo} and probe a wide range of AGN luminosities (\LbolEq{=}{43-47}) and ISM substructures.
In homogeneous discs, the AGN wind sweeps up an outflowing, cooling shell, where the emerging cold phase dominates the mass and kinetic energy budgets, reaching a momentum flux $\dot{p} \approx 7\ L/c$.
However, when the ISM is clumpy, outflow properties are profoundly different. They contain small, long-lived ($\gtrsim 5\ \rm{Myr}$), cold ({\TEq{\lesssim}{4.5}}) cloudlets entrained in the faster, hot outflow phase, which are only present in the outflow if radiative cooling is included in the simulation. While the cold phase dominates the mass of the outflow, most of the kinetic luminosity is now carried by a tenuous, hot phase with $T \gtrsim 10^7 \, \rm K$. While the hot phases reaches momentum fluxes $\dot{p} \approx (1 - 5)\ L/c$, energy-driven bubbles couple to the cold phase inefficiently, producing modest momentum fluxes $\dot{p} \lesssim  L/c$ in the fast-outflowing cold gas.
These low momentum fluxes could lead to the outflows being misclassified as momentum-driven using common observational diagnostics.
We also show predictions for scaling relations between outflow properties and AGN luminosity and discuss the challenges in constraining outflow driving mechanisms and kinetic coupling efficiencies using observed quantities.

\end{abstract}




\section{Introduction}
\label{sec:c4_intro}

At the centres of massive galaxies lie supermassive black holes \citep{Kormendy2013}. These objects have masses ranging from $\sim 10^6-10^{10} \rm{\ M_{\odot}}$, and during periods of intense gas accretion can `light up' to become Active Galactic Nuclei (AGN). Over the lifetime of a supermassive black hole, more than the binding energy of the galaxy can be released, giving AGN the potential to influence the fate of their host galaxies; a process known as AGN feedback \citep[e.g.][]{Fabian2012, King2015}. AGN feedback is an essential component in all contemporary theoretical models and simulations of galaxy formation \citep[e.g.,][]{Hirschmann2014,Schaye2015,Weinberger2018,Dave2019,Dubois2021,Wellons2023}. However, direct observational evidence of this feedback on a population scale is lacking, with studies finding that AGN-hosting galaxies are no more quenched or gas-depleted than their inactive counterparts (\citealt{Rosario2013a,Rosario2018,Kirkpatrick2019,Jarvis2020,Shangguan2020a,Valentino2021,Koss2021,Zhuang2021,Ji2022,Kim2022,FriasCastillo2024}, but see some counterexamples at higher redshift: \citealt{Perna2018,Bischetti2021,Circosta2021,Bertola2024}). As shown in previous work (\citealt{Ward2022}, see also \citealt{Scholtz2018,Piotrowska2022}), this apparent tension is not in contradiction with models that rely on AGN feedback to quench galaxies. It instead highlights the difficulty in studying this problem due to the vast range in timescales and distances involved \citep{Harrison2017}, and motivates further work on the impact of AGN feedback on the host galaxy.

The power of an AGN can be coupled through jets \citep[e.g.][]{Mukherjee2016,Bourne2023}, accretion disc winds \citep{Silk1998,King2003,Faucher-Giguere2012,Almeida2023} and/or direct radiation pressure \citep{Thompson2015,Ishibashi2018,Costa2018a}, all of which can drive kpc-scale outflows in the host galaxy, transferring mass, momentum and energy to the interstellar medium (ISM) and circumgalactic medium (CGM). Studying the interaction between these AGN-driven outflows and the multiphase ISM in galaxies is therefore critical in understanding the role AGN play in galaxy evolution \citep{Harrison2024}.
To this end, the last decade has seen an explosion of observational work studying the multiphase nature of AGN outflows. For example, radio and sub-mm observatories have allowed measurements of the coldest phase of the gas (often using CO, [{\ion{C}{II}}] or \ion{H}{I} transitions; e.g., \citealt{Morganti2005,Cicone2014,Fluetsch2019,Veilleux2020,Lamperti2022,RamosAlmeida2022,Girdhar2024}); and X-ray--NIR spectroscopic data has enabled measurements of high velocity disk winds \citep[e.g.,][]{King2015,Gofford2015,Chartas2021,Matzeu2023}, the ionised or atomic phases in the ISM (e.g., using [\ion{O}{III}] or Na I D; \citealt{Rupke2013,Harrison2014,Zakamska2014,Molyneux2019,Musiimenta2023}), and X-ray imaging/spectroscopic studies for the hottest gas on large scales \citep[e.g.,][]{Lansbury2018,Longinotti2023}. Such studies often seek to quantify the mass outflow rate, momentum rate and kinetic coupling efficiency of the outflows. These quantities are often used to infer if the outflows are momentum- or energy-driven, and if the outflows are considered energetic enough to impact the host \citep[e.g.,][]{Cicone2014,Lamperti2022,Riffel2023}. 

Some studies have identified scaling relations between outflow properties, such as the mass outflow rate or kinetic energy coupling efficiency, and the AGN luminosity \citep[e.g.,][]{Fiore2017,Leung2019,Davies2020,Musiimenta2023}. However, these trends may partly be driven by selection effects, such as the choice to target CO-bright systems with known outflows in other phases (see discussion in \citealt{RamosAlmeida2022,Harrison2024}). These efforts are further hampered by observational challenges; for example, the spatial extent of the outflow can be challenging to measure, especially if the resolution of the observations is only marginally better than the galaxy size; the velocity of the outflow is difficult to disentangle due to systemic galaxy motion and projection effects; and poorly constrained conversion factors and electron densities add large uncertainties to the outflow mass estimates \citep{Husemann2016,Tadhunter2018,Harrison2018,Veilleux2020}. Furthermore, if only one gas phase is measured, a significant amount of the outflow mass or energy will be missed, potentially changing the resulting conclusion on the outflow driving mechanism. Therefore, an unbiased, multiwavelength approach is essential \citep{Cicone2018}.  

To help interpret the results of these observations, comparisons to predictions from simulations need to be made \citep[e.g.,][]{Meenakshi2022}. In particular, an outstanding task is to establish the theoretical expectations of scaling relations and examine how these compare to the observed trends. However, accurately simulating the interaction between the ISM and AGN-driven outflows is a complex numerical task as the dynamic range of the problem is vast, from outflows than can reach 100s kpc to sub-pc-scale structures in the ISM. One of the major challenges is self-consistently modelling realistic ISM conditions. The complex structure of the ISM is maintained by a range of physical processes, such as supernova and star formation feedback \citep{Gent2013}, dust formation and destruction \citep{Hirashita2020,Kirchschlager2022}, molecular chemistry and cooling \citep{Richings2018a,Richings2018b}, and cosmic ray feedback \citep{Ruszkowski2023}. Simulating all these effects, especially on a galaxy scale, is computationally challenging, both in terms of the complex physical processes involved and the resolution requirements. Therefore, most galaxy and cosmological-scale simulations use an effective equation of state model (eEOS; e.g., \citealt{Springel2003}). This packages up the unresolved ISM physics such as star formation and feedback into a subgrid model, allowing galaxies to be simulated without resolving all the physical processes involved \citep[e.g.,][]{Vogelsberger2013}. However, this simplified model does not accurately reproduce the spatial or multiphase structure of the ISM, which may play a significant role in the propagation of outflows through the galaxy. To overcome the inability to resolve the ISM substructure, some studies manually set the spatial distribution of the gas, creating a clumpy, two-phase media composed of fractally-distributed cold gas clouds surrounded by a tenuous hot phase. This setup also allows the effect of wind-ISM interactions to be studied in inhomogeneous environments, unlike analytic models, which often assume spherically-symmetric media, and eEOS models, which do not capture small-scale ISM structure. This method has been used by previous studies to investigate the effect of jets \citep{Wagner2011,Mukherjee2016,Tanner2022}, winds from AGN \citep{Wagner2013} and starbursts \citep{Cooper2008}, radiation pressure \citep{Bieri2017}, and shocks \citep{BandaBarragan2020,BandaBarragan2021} in clumpy media.

Another open question is how cold gas becomes entrained in a hot outflow. The crushing time of cold clouds is shorter than their entrainment timescale \citep{Klein1994,Zhang2017}. Yet cold gas clouds are observed in outflows travelling at $100 \rm{s\ km\  s^{-1}}$ \citep{DiTeodoro2019,Veilleux2020}. Recent simulations of the interaction between single clouds and a hot wind have shown that the clouds can survive, and even increase in mass, due to efficient radiative cooling at the mixing boundary between the cloud and the wind \citep{Gronke2018,Gronke2020a,Fielding2020}, although this effect is also sensitive to the density structure of the clouds themselves \citep{BandaBarragan2019,Mandal2024}. Therefore, understanding how these small clouds form, and resolving their interaction with the hot wind, is crucial in understanding the cold outflows seen in observations.

In this work, we set up a series of computational experiments to study the interaction between a clumpy ISM and a fast, small-scale AGN accretion-disc wind, modelled following \cite{Costa2020}. This paper is the first in our new project "AGN in Clumpy DisCs" (ACDC). The main questions we set out to answer are:

\begin{enumerate}
    \item What are the properties of outflows formed from wind-ISM interactions?
    \item How does having a clumpy ISM affect the properties of multiphase outflows compared to a smooth medium?
    \item How can observed outflow measurements, such as momentum rate and kinetic energy coupling efficiency, constrain AGN feedback mechanisms?
    \item Do we expect scaling relations between outflow properties and AGN luminosity?
\end{enumerate}

We structure our paper as follows: in Section \ref{sec:method} we present our experimental setup, including our method for creating a clumpy ISM (Section \ref{sec:ISM_setup}) and our AGN wind model (Section \ref{sec:BOLA}). We split our results into two parts; in Section \ref{sec:formation} we show the formation of multiphase outflows in our simulations and quantify their properties. In Section \ref{sec:implications} we then discuss the implications of our findings for observational studies. We finish by presenting our conclusions and discussing directions for future studies in Section \ref{sec:outlook}.

\begin{figure*}
    \centering
    \includegraphics[width=\textwidth]{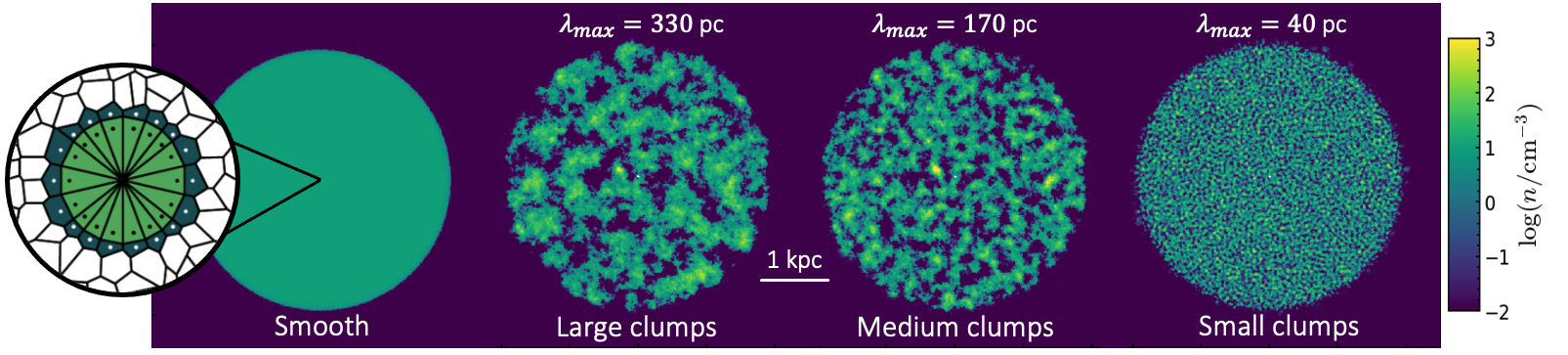}
    \caption{The initial conditions for our different simulations, showing a face-on slice through our disc. We explore a range of largest cloud sizes of \LmaxEq{=}{40}, \LmaxEq{=}{170} and \LmaxEq{=}{330}, and a smooth disc for comparison. The disc gas mass is $M_{\rm{disc}}=1.4{\times}10^9 M_\odot$ and the ambient background gas is in pressure equilibrium with the clumps, and has a temperature of \TEq{=}{7}. Presented on the left is a schematic of the \textsc{Bola} boundary structure \citep{Costa2020}, showing the two layers of \textsc{Arepo} cells used to launch the AGN wind.}
    \label{fig:method}
\end{figure*}

\section{Numerical Simulations} \label{sec:method}

To investigate the effect of a clumpy ISM on the propagation of AGN wind-driven outflows, we set up a series of controlled experiments. These feature an idealised galaxy disc within a uniform gaseous halo where the structure of gas in the disc has been manually set to mimic the observed fractal substructure of the ISM (similar to \citealt{Wagner2011,Wagner2012,Wagner2013,Mukherjee2016,Bieri2017,Mukherjee2018,Tanner2022}). An AGN is placed at the centre of the disc and a spherical wind solution is set up via \textsc{Bola} (\citealt{Costa2020}; see Section \ref{sec:BOLA}). Using this setup, we can investigate the interaction between an AGN wind and a spatially resolved ISM. By varying the ISM structure and the properties of the AGN wind, we can also evaluate how these affect the resulting outflow. This is motivated both by previous numerical studies, which found that the initial sizes of clouds in the ISM was a critical parameter in determining feedback efficiency \citep{Wagner2012}, and by observations which suggest that outflow properties correlate strongly with AGN luminosity \citep[e.g.,][]{Fiore2017}.

In Section \ref{sec:hydro} we present the hydrodynamic code used, before explaining how the galaxy disc (Section \ref{sec:galaxy_setup}), ISM structure (Section \ref{sec:ISM_setup}) and AGN wind (Section \ref{sec:BOLA}) were set up. Finally, we describe the suite of simulations explored in Section \ref{sec:runs} and we explain the calculations used to extract outflow properties in Section \ref{sec:method_outflow}.

\subsection{Hydrodynamic code} \label{sec:hydro}

Our simulations are performed using the hydrodynamic code \textsc{Arepo} \citep{Springel2010}. This uses an unstructured, Voronoi mesh that moves, in a quasi-Lagrangian fashion, with the fluid. This maintains Galilean invariance whilst also providing excellent shock capturing without the need for artificial viscosity, as in particle-based codes. Primitive variables are linearly reconstructed within each cell, providing second-order spatial accuracy. Extrapolated values at cell boundaries are used to compute hydrodynamic fluxes using an exact Riemann solver \citep{Pakmor2016}. Cells are refined or de-refined according to a pre-determined refinement criterion. By default, this involves (de)refining to keep cells within a factor of two from a target mass. This results in high spatial resolution in regions of high density, allowing us to resolve small-scale structures such as the clumpy distribution of the ISM on pc-scales. For our fiducial-resolution simulations, we use a target gas mass of $M_{\rm{target}}=100\ M_\odot$ which gives us a maximum spatial resolution of around $d_{\rm{cell}}=1\rm{\ pc}$ (see Figure \ref{fig:app_cell}). We discuss the numerical convergence of our results in Appendix \ref{ap_resolution}. We also extend the refinement criteria to increase the resolution in the AGN wind (see Section \ref{sec:BOLA}).

\subsection{Galaxy setup} \label{sec:galaxy_setup}

In this study, we analyse an idealised galaxy disc located within a hot halo. We use a box size of $L_{\rm{box}}=20\ \rm{kpc}$ with periodic boundary conditions. The simulations are performed for a period of time shorter than the outflow crossing time (\tEq{=}{5} for our fiducial simulations). The disc has a diameter of 4 kpc and a thickness of 1 kpc (following similar setups in \citealt{Mukherjee2018,Tanner2022}) and tapers at the edges to smooth the interface between the background. The mean gas number density in the disc is set to $\langle n_0 \rangle = 5\rm{ cm^{-3}}$, resulting in a disc gas mass of $M_{\rm{disc}}=1.4{\times}10^9 \rm{\ M_\odot}$. The background is set in pressure equilibrium with the disc, giving temperatures of $T_{0,\rm{disc}}=10^4\ \rm{K}$ and $T_{0,\rm{bkg}}=10^7\ \rm{K}$ and a number density of $n_{\rm{bkg}}=10^{-2}\ \rm{cm^{-3}}$. In our fiducial simulations, we also include the standard \textsc{Arepo} prescription for primordial cooling, excluding a UV background. We investigate the effect of cooling in Section \ref{sec:energetics}. We also investigate the effect of altering the disc mass by reducing the disc height (to $h=0.5 \rm{\ kpc}$) and reducing the mean gas density (to $\langle n_0 \rangle = 2.5\rm{\ cm^{-3}}$). The sensitivity of our results to these changes is explored in Section \ref{sec:sensitivity}.

\subsection{Setting the ISM structure} \label{sec:ISM_setup}

In this study, we take a controlled experiment approach to investigating the interaction between an AGN wind an clumpy ISM. We manually create a two-phase ISM of cold clumps, arranged in a fractal distribution, surrounded by a hot, diffuse phase. Thanks to our high  resolution, this allows us to spatially resolve the ISM structure to investigate what effect this has on the AGN wind moving through the galaxy, and vice-versa.

To initialise our ISM structure, we make use of the PyFC\footnote{\url{https://pypi.org/project/pyFC/}} Python package \citep{Wagner2012}. This generates a random, three-dimensional scalar field from a given probability distribution function with a fractal spatial correlation, a method introduced by \cite{Sutherland2007} based on a scheme developed for terrestrial clouds by \cite{LewisAustin2002}. We use a log-normal distribution function, with width $\sigma=\sqrt{5}$, and a Kolmogorov power-law spectrum ($\beta = - \frac{5}{3}$), motivated by observed ISM density distributions \citep{Fischera2003}. The resulting fractal structure is parameterised by the lower wavenumber cutoff, $k_{\rm{min}}$, which represents the largest correlated spatial scale. This can be related to the average largest cloud size, $\lambda_{\rm{max}}$, by
\begin{equation}
    \lambda_{\rm{max}} = \frac{L}{2 k_{\rm{min}}} \, ,
\end{equation}
where $L$ is the box length of the fractal cube being generated. Low values of $k_{\rm{min}}$ represent large initial cloud sizes, and high values represent small initial cloud sizes. The maximum value of {\kmin} that can be used is set by the Nyquist limit which depends on the simulation resolution. In our study, we use a range of minimum wavenumber values, from $6 \rm{\ kpc^{-1}}$ $\leq {k}_{\rm{min}} \leq 50 \rm{\ kpc^{-1}}$ which corresponds to average largest cloud sizes of $40 \rm{\ pc} \leq \lambda_{\rm{max}} \leq 333 \rm{\ pc}$, giving a wide range of ISM conditions. The resulting fractal cube is then cropped into the desired disc shape and the densities scaled to the mean disc density. To generate porosity in the ISM, cells above a temperature threshold of $T_{\rm{crit}}=3\times 10 ^4\ \rm{K}$ are considered thermally unstable and replaced by gas of the same temperature as the background, in pressure equilibrium with the cold clumps \citep[following][]{Sutherland2007, Cooper2008}. This generates a two-phase medium of fractally-distributed cold clumps and hot, diffuse gas. The edges of the disc ($>400 \rm{\ pc}$ above the midplane) are then tapered with a $tanh$ profile \citep{Tanner2022} to create a smoother boundary with the background. We note that using this method maintains a constant disc mass regardless of the value of {\kmin}.

As this method was designed for grid-based (constant cell volume) codes, a further step is required to set-up the disc for use with \textsc{Arepo} which uses the constant mass approach. The Voronoi grid is first evolved with hydro fluxes disabled, which allows the grid to regularise \citep{Springel2010} and to refine (de-refine) in regions of high (low) density. Once the total number of cells has converged, the disc retains the desired fractal properties, but now with a quasi-Lagrangian setup. The resulting initial conditions are shown in Figure \ref{fig:method} for a range of clumping factors, alongside our smooth disc case. We can see that the initial clumps have a range of sizes and densities, from $n\approx 10^{1-3} \rm{\ cm^{-3}}$, roughly in line with what is expected for cold \ion{H}{I} clouds \citep{Cox2005}.

The advantage of using this manually-set ISM structure is it allows us to conduct controlled experiments of the interaction between an outflow and a spatially-resolved ISM without attempting to create an ISM structure from first-principles, which would be highly dependent on a range of loosely-constrained subgrid models. It also allows very high spatial resolution to be achieved to study the small-scale structure of the ISM and its resulting fragmentation. However, this method means our simulations are very idealised and only run for short timescales ($\leq 5 \rm{\ Myr}$), both due to boxsize constraints and cooling losses, which cause the fractal structure to diffuse on longer timescales. We also neglect effects such as self-gravity, turbulence and star formation. We note that some studies using this method do include a static gravitational potential \citep[e.g.,][]{Mukherjee2016,Tanner2022} and an initial gas velocity dispersion \citep{Mukherjee2018}. This has the effect of changing the initial clumps to a filamentary structure and smoothing out the density contrast between the clouds and the porous gaps to create conditions more similar to a turbulent medium. However, this makes it harder to control the porosity of the ISM, so we follow \cite{Wagner2012, Bieri2017} and neglect gravity to focus solely on the interaction between the quasar wind and our manually-set ISM structure. We note that the outflow crossing time of the fastest-moving gas is shorter than the free-fall time of the galaxy ($\approx 50\ \rm{Myr}$) meaning it is unlikely to be significantly affected by gravity \citep[see][]{Wagner2012}. However, the slower-moving tail of the outflow is limit.

\subsection{AGN wind model} \label{sec:BOLA}

To model an AGN wind, we employ the \textsc{Bola} (BOundary Layer for AGN) model introduced in \cite{Costa2020}. A sphere of cells are fixed in place at the location of the black hole. This is composed of two layers: an `inner layer', which is excluded from the hydrodynamic calculations and used only to define the fixed boundary, and an `outer layer', which is free to evolve hydrodynamically. Both layers are discretised following a \textsc{Healpix} tessellation into $12n_{\rm{side}}^2$ pixels of equal surface area  \citep{Gorski2005}. In this study, we use a value of $n_{\rm{side}}=8$. We set the radius of the inner spherical boundary as $r_{\rm{sp}}=10 \rm{\ pc}$. The mass, momentum and energy flux is set at the interface between the two layers and then communicated from the outer layer into the surrounding gas. \textsc{Bola} makes it possible to model feedback processes operating below the resolution scale through appropriate choice of boundary conditions. Here, we adopt boundary conditions for a spherical, ultra-fast outflow (UFO) as in \citet{Costa2020}. These boundary conditions directly reproduce the UFO conditions investigated in detail in \cite{King2003, Faucher-Giguere2012, Costa2014} which lead to energy-driven bubbles that can produce strong large-scale feedback. In addition, a passive scalar is injected across the boundary which is then advected along with the injected wind. This allows \textsc{Bola} to modify the \textsc{Arepo} refinement scheme to increase the resolution of wind cells by boosting the resolution in cells with a high density of wind fluid. This helps to reduce the problem that the wind itself is poorly-resolved due to its low density. We decrease the target mass of wind cells by a factor of 10.

The main free parameters for this model are the AGN luminosity ({\LAGN}), wind velocity ({\vAGN}), temperature ($T_{\rm{AGN}}$) and momentum boost factor $\tau = \dot{p}/(L_{\rm{AGN}}/c)$ (all at injection scale). We use a momentum boost factor of $\tau=1$ and an initial wind temperature of $T_{\rm{AGN}}=10^6 \rm{\ K}$. These parameters ensure that the wind remains highly supersonic out to the free-expansion radius, and that the pressure contribution to the kinetic luminosity and momentum flux is marginal (see \citealt{King2003,Faucher-Giguere2012}). Furthermore, the choice of initial wind temperature plays a negligible role, as the wind is quickly shocked to higher temperatures. We take fiducial values for the AGN luminosity and wind velocity of {\LbolEq{=}{45}} and {\vAGNEq{=}{10,000}} but explore a range of parameters in our simulation, as shown in Table \ref{table:runs}. The initial kinetic luminosity injected is given by
\begin{equation}
    \dot{E}_w = \frac{\tau \beta}{2} L_{\rm{AGN}} \, ,
    \label{eq:E_k_init}
\end{equation}
where $\beta=v_{\rm{AGN}}/c$. For our chosen values of $\tau=1$ and \vAGNEq{=}{10,000}, this gives an initial energy coupling efficiency of $\dot{E}_w/L_{\rm AGN} = 1.7\%$, in the fiducial case. These values are consistent with studies of ultra-fast outflows detected at small-scales in X-rays \citep[e.g.,][]{Gofford2015,Matzeu2023}. It is important to note that these injected values are on the scale of the AGN accretion disc (sub-pc) and are therefore not necessarily comparable to the resulting large-scale outflows on kpc-scales \citep{Harrison2018,Costa2020}. We discuss the derived momentum boost and kinetic coupling efficiency of the large-scale outflow in Section \ref{sec:challenges_driving}. In this study we don't explore gas accretion onto the AGN, and our AGN maintains a constant luminosity throughout.

\begin{table*}
\begin{tabular}{@{}llrrrrrrrr@{}}
\toprule
Simulation Name                  & {\Mtarget} & {\kmin}  & {\Lmax} & $h$   & $\langle n_0 \rangle$ & Cooling    & {\LAGN} & {\vAGN} \\ 
 &                  [$M_\odot$] & [$\rm{kpc}^{-1}$]  & [$\rm{pc}$] & [$\rm{kpc}$]    & [$\rm{cm^{-3}}$] &    & [$\rm{erg\ s^{-1}}$]  & [$\rm{km\ s^{-1}}$] \\ \midrule

Small clumps &  100             & 50               & 40              & 1          & 5                              & Y          & $10^{45}$         & 10000          \\
\textbf{Medium clumps (Fiducial)} &  \textbf{100}    & \textbf{12}      & \textbf{170}    & \textbf{1} & \textbf{5}                     & \textbf{Y} & $\mathbf{10^{45}}$ & \textbf{10000} \\
Large clumps &  100             & 6                & 330             & 1          & 5                              & Y          & $10^{45}$         & 10000          \\
Smooth &  100             & - & -               & 1          & 5                              & Y          & $10^{46}$         & 10000          \\
No cooling &  100             & 12               & 170             & 1          & 5                              & N          & $10^{45}$        & 10000          \\
 Slow wind&  100             & 12               & 170             & 1          & 5                              & Y          & $10^{45}$          & 5000           \\
 Thin disc &  100             & 12               & 170             & 0.5        & 5                              & Y          & $10^{45}$          & 10000          \\
 Low density &  100             & 12               & 170             & 1          & 2.5                               & Y          & $10^{46}$          & 10000          \\
 
 L43 &            100             & 12               & 170             & 1          & 5                              & Y          & $10^{43}$         & 10000          \\
 L44&  100             & 12               & 170             & 1          & 5                              & Y          & $10^{44}$         & 10000          \\
 L46&  100             & 12               & 170             & 1          & 5                              & Y          & $10^{46}$        & 10000          \\
 L47&  100             & 12               & 170             & 1          & 5                              & Y          & $10^{47}$         & 10000          \\

 No AGN &  100             & 12               & 170             & 1          & 5                              & Y          & -                               & -              \\
Low-res &  1000            & 12               & 170             & 1          & 5                              & Y          & $10^{45}$          & 10000          \\
 Low-res, no cooling&  1000            & 12               & 170             & 1          & 5                              & N          & $10^{45}$          & 10000          \\
 High-res &  10            & 12               & 170             & 1          & 5                              & Y          & $10^{45}$          & 10000          \\ \bottomrule
\end{tabular}
\caption{The suite of simulations explored in this work. In bold is our fiducial simulation. The columns represent: 1) The nickname for each run; 2) The target mass resolution; 3) The wavenumber of the initial clump sizes; 4) The average maximum cloud size of the clumps; 5) The height of the disc; 6) The initial mean number density of the clumps; 7) Whether radiative cooling is included; 8) The luminosity of the AGN; and 9) The velocity of the AGN wind at injection.}
\label{table:runs}
\end{table*}

\subsection{Simulation suite} \label{sec:runs}

Table \ref{table:runs} shows the range of simulation parameters investigated in this study. For clarity, we name the ones most commonly discussed in the paper. In particular, we highlight `Medium clumps' as our fiducial simulation: this has a clumping parameter of {\kminEq{=}{12}} (\LmaxEq{=}{170}), an AGN luminosity of {\LbolEq{=}{45}} and includes radiative cooling. The simulation with the same parameters, but without cooling, is called `No Cooling'. For comparison, we also analyse the resulting outflow from an AGN place in a homogeneous disc, with a constant density of $n_0 = 5 \ \rm{cm^{-3}}$ (`Smooth'). We show the results for the smooth disc in Section \ref{sec:smooth_disc_qual} and compare the phase content and energetics of the resulting outflow to the cooling and no-cooling cases in Section \ref{sec:energetics}. In Section \ref{sec:kmin_vr}, we investigate the effect of increasing (`Large clumps': {\LmaxEq{=}{330}}) and decreasing (`Small clumps': {\LmaxEq{=}{40}}) the average sizes of the initial clouds (see Figure \ref{fig:method}). In Section \ref{sec:sensitivity} we investigate the sensitivity of our results to varying the initial parameters, in particular, by reducing the height of the disc (`Thin disc'), decreasing the initial mean clump density (`Low density') and reducing the initial AGN wind speed (`Slow wind'). In Section \ref{sec:fiore}, we investigate a range of AGN luminosities (\LbolEq{=}{43-47}) to ascertain if we predict scaling relations between the luminosity, and the outflow mass or kinetic energy coupling efficiency. Finally, we also simulate our fiducial setup at lower (\MtargetEq{=}{1000}) and higher (\MtargetEq{=}{10}) resolutions to investigate the convergence of our simulations, which we show in Appendix \ref{ap_resolution}.

\subsection{Calculating outflow properties} \label{sec:method_outflow}

\begin{figure*}
   \centering
    \includegraphics[width=0.88\textwidth]{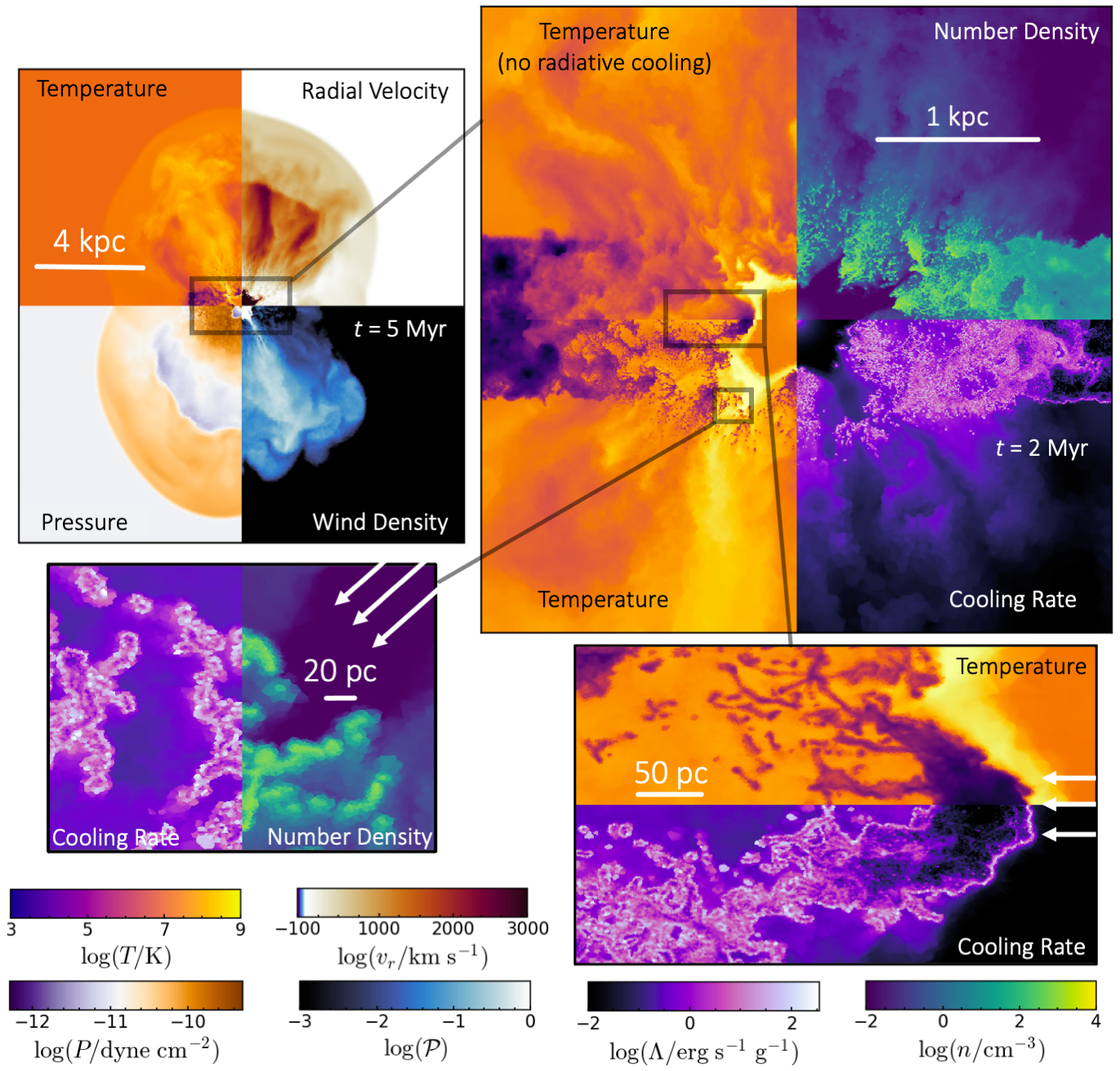}
    \caption{Summary figure showing the results of our fiducial simulation at a range of spatial scales, as detailed in Section \ref{sec:multiscale}. We show our fiducial model, with medium-sized clumps ({\LmaxEq{=}{170}}), an AGN luminosity of {\LbolEq{=}{45}} and a wind injection velocity of {\vAGNEq{=}{10,000}}. The top left panel shows the large-scale biconical outflow propagating into the halo at {\tEq{=}{5}} and the top right panel shows the central region of the disc at \tEq{=}{2}, showing the AGN wind structure and the effect of neglecting radiative cooling (top left quadrant). The bottom two panels show the small scale fragmentation and entrainment of cold gas at {\tEq{=}{2}} at different spatial scales. Scale bars are shown for reference in each panel, and the white arrows demonstrate the direction of the AGN wind in the bottom two panels. All plots share the same colour scales, with corresponding values shown by the colourbars at the bottom.}
    \label{fig:summary}
\end{figure*}

A key aspect of this work is to evaluate the mass $M_{\rm{OF}}$, outflow rate $\dot{M}_{\rm{OF}}$, momentum flux $\dot{p}$ and kinetic luminosity $\dot{E}_k$ of the outflow, which are commonly used in observational studies to characterise AGN-driven outflows \citep[e.g.,][]{Cicone2014,Fiore2017,Gonzalez-Alfonso2017,Bischetti2019,Musiimenta2023}. There are a diversity of approaches used in observational studies to derive these values \citep[see reviews in][]{Harrison2018,Veilleux2020,Harrison2024}, which we will discuss in Section \ref{sec:challenges_measure}. As a baseline method, we take a time-averaged approach to calculate these global quantities from our simulations. Additionally, calculating the mass contained in the outflow is challenging observationally and most approaches have some sensitivity to the minimum velocity of gas that can be detected as part of the outflow (see discussion in Section \ref{sec:outflow_v}). Therefore, to emulate the limitation of observations, we use a minimum radial velocity cut ({\vmin}) to mimic isolating the outflow gas from that attributed to host galaxy dynamics. Because our simulations have an idealised, initially isobaric setup with no rotation or velocity dispersion, any gas that is moving should be due to the AGN wind. In practice, there is some slight collapse of the initial clouds due to cooling, but selecting a definition of {\vminEq{=}{10}} yields a contamination rate of $<1\%$ based on a comparison to our non-AGN simulation. This value is also of the same order as the sound speed in cold clouds ($c_s \approx 15 \rm{km/s}$). Therefore we take a `theoretical' cut of {\vminEq{=}{10}} to represent the outflow directly caused by the AGN. However, in an observed galaxy, gas will also be in motion due to rotation, turbulence, star formation-driven outflows, etc. Such a low cut of {\vminEq{=}{10}} for an AGN-driven outflow may often not be clearly separable from other gas motions within the host galaxy \citep[e.g.,][]{Marconcini2023,Harrison2024}. Therefore, we also investigate a cut of {\vminEq{=}{100}} as a representative example of a more `observational' limit. The effect of {\vmin} on our results is discussed in Section {\ref{sec:outflow_v}}.

To calculate the global, time-averaged outflow mass, we take a cell-by-cell sum of all the gas moving at radial velocities $>v_{\rm{min}}$,
\begin{equation}
    {M}_{\rm{OF}} = \sum_{i}^{\rm{for\ } v_i > v_{\rm{min}}} m_i \, .
    \label{eq:S2_Mdot}
\end{equation}
Mass outflow rates are then calculated as time-averages by dividing by the simulation time ($t_{\rm{sim}}$),
\begin{equation}
    \dot{M}_{\rm{OF}} = \frac{1}{t_{\rm{sim}}}   \sum_{i}^{\rm{for\ } v_i > v_{\rm{min}}} m_i \, ,
\end{equation}
while for the momentum rates, this is 
\begin{equation}
    \dot{p}_{\rm{OF}} = \frac{1}{t_{\rm{sim}}}  \sum_{i}^{\rm{for\ } v_i > v_{\rm{min}}} m_i v_i \, ,
    \label{eq:S2_mom}
\end{equation}
and the kinetic energy flux reads
\begin{equation}
    \dot{E}_{\rm{OF}} = \frac{1}{2 t_{\rm{sim}}}  \sum_{i}^{\rm{for\ } v_i > v_{\rm{min}}} m_i  v_i^2 \, .
    \label{eq:S2_E}
\end{equation}

We apply these equations to the simulations presented in Section \ref{sec:runs} to investigate the outflow properties of an AGN wind interacting with a clumpy ISM structure.

\section{Results I: Multiphase outflows from small-scale winds} \label{sec:formation}

In this section, we present the results of our simulations of an AGN wind in an idealised, clumpy galaxy disc. We first introduce our fiducial simulation, showing the multiscale structure of the outflows generated (Section \ref{sec:multiscale}), before analysing the phase structure and energetics (Section \ref{sec:energetics}). We then discuss how changing the parameters of our galaxy setup affects our results (Section \ref{sec:variation}).

\subsection{Multiscale outflow structure} \label{sec:multiscale}

Figure \ref{fig:summary} shows a qualitative overview of our simulations, showing the formation of a multiscale, multiphase outflow caused by the AGN wind. We use our fiducial simulation, with intermediate initial clumpiness\footnote{Note: we will use the term `clumps' to refer to initial overdensities in the density field, and the term `cloud' or `cloudlet' to refer to the resulting fragments entrained in the outflow} (corresponding to a largest clump size of {\LmaxEq{=}{170}}) and mass resolution {\MtargetEq{=}{100}}. The AGN has a luminosity of {\LAGNEq{=}{45}} and an initial wind velocity of {\vAGNEq{=}{10\ 000}}. We show the large-scale structure of the outflow in the top left at {\tEq{=}{5}}, showing the maximum outflow extent reached in the simulation. The other panels show the galaxy at {\tEq{=}{2}}, to demonstrate the interactions between the wind and the initial cold clumps.

\begin{figure}
    \centering
    \includegraphics[width=0.45\textwidth]{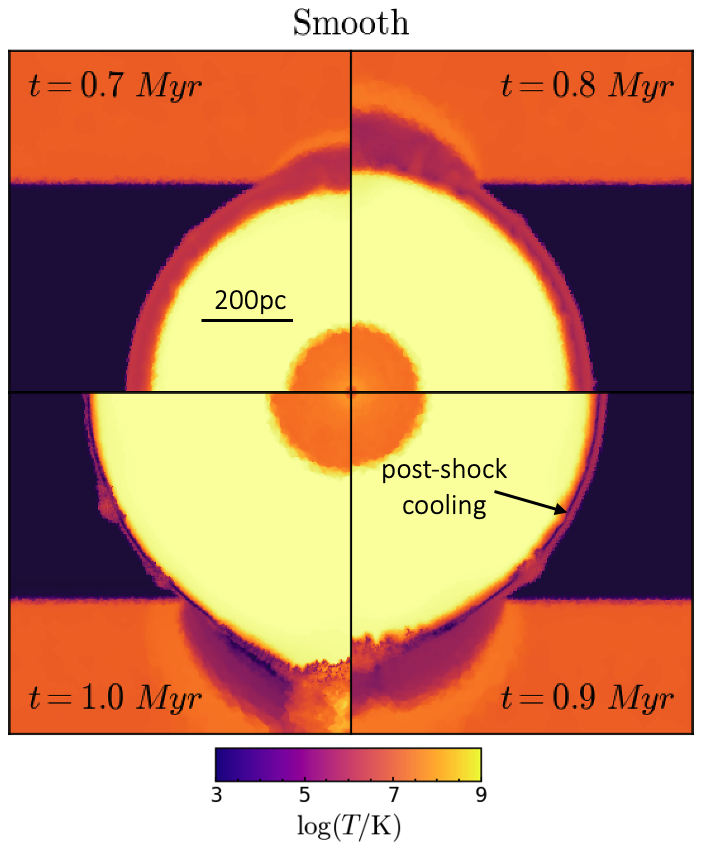}
    \caption{Our simulation with an initially smooth disc. Time increases clockwise from the top-left. The AGN wind sweeps up material forming a thin shell of outflowing material. Once this reaches a critical density, it cools suddenly via post-shock cooling, resulting in a cold (\TEq{<}{4.5}) outflow at \tEq{\geq}{0.8}. The resulting morphology and energetics are starkly different from a wind launched in an inhomogeneous medium.}
    \label{fig:smoothdisc}
\end{figure}

\textbf{Halo scale:} In the top left panel, we show an edge-on slice of the galaxy and halo, presenting (clockwise  from top left quadrant) the temperature, radial velocity, wind tracer density and pressure. At this large scale, the outflow is biconical, with a clear forward shock propagating into the halo at a radius of around 6\,kpc. Although the wind is injected spherically on small scales, it encounters more mass and slows down to lower velocities in the equatorial direction, leading to a bipolar outflow emerging from the top and bottom of the disc \citep[e.g.,][]{Costa2014,Nelson2019b}. The outskirts of the region populated by the wind tracer (lower-right quadrant) correspond to a contact surface separating shocked wind and shocked ambient medium fluid \citep{Costa2020}. Due to the initially inhomogeneous disc medium, we also see several high velocity ({\vrEq{\gtrsim}{2000}}) `chimneys' where the wind escapes the quickest. This can be most clearly seen in the radial velocity and wind tracer panels as narrow streams of the fastest-moving gas which create elongated structures a few kpc long, before running into the back of the shocked ambient medium at $4 \rm{\ kpc}$. The overall structure of the wind-driven outflow is in line with previous theoretical work (e.g., \citealt{King2003,Faucher-Giguere2012,Zubovas2012,Costa2014,Nims2015,Costa2020}, see also our Figure \ref{fig:smoothdisc}) with the addition of the high-velocity chimneys caused by the initial clumpiness.

\textbf{Disc scale:} The top right panel of Figure \ref{fig:summary} shows a closer view of the disc itself, showing (clockwise from top left quadrant) the temperature (in a simulation without radiative cooling) then the number density, cooling rate, and temperature, all in the fiducial simulation including cooling. In the centre, we can see the freely-expanding wind, and the forward and reverse shock fronts. We also see clear bow shocks forming around the initial clumps and free streaming of the wind in low-density regions, leading to the aforementioned `chimneys'. In the left-hand half of the panel, we compare the effect of including and excluding radiative cooling in the simulations. The non-cooling simulation (top left quadrant) shows a more continuous billowing of gas as it is blown away by the wind. The outflowing gas is hotter than in the cooling case, resulting in very little cold (\TEq{<}{4.5}) outflowing gas. However, in the cooling case (bottom left quadrant in the top right panel), we see that the clumps cool, collapse and fragment into much smaller cloudlets than the initial clump size. These clouds become entrained in the hotter, faster wind, creating an outflow of relatively cold (\TEq{<}{4.5}) gas contained in many small, high density cloudlets. We note that the exact size of these fragments is likely resolution-dependent; we explore the numerical convergence of our simulations in Appendix \ref{ap_resolution}.

\textbf{Cloud scale:} The bottom two panels of Figure \ref{fig:summary} show the cloudlet features in more detail. The bottom right panel shows the temperature (top half) and cooling rate per unit mass (bottom half) of one high-density clump. The AGN wind approaches from the right (white arrows), creating a strong bow shock around the clump and creating a tail of cold fragments and filaments \citep[see also][]{Cooper2008}. The mixing layer between the cold and the hot gas cools rapidly. This cooling `skin' around the clouds very likely leads to cold cloud formation in the outflow \citep{Gronke2018,Schneider2018, Gronke2020a, Fielding2020}, although a full exploration of this is left to future work. The bottom left panel of Figure {\ref{fig:summary}} shows some of the entrained cloudlets, with the cooling rate on the left and the number density on the right. These cloudlets are small ($d_{\rm{cloud}} \approx 10 - 20 \, \rm pc$), dense ($n \approx 10^3 \rm{cm^{-3}}$; generally more dense than the initial clumps), and rapidly cooling, especially at their surface. The cooling time for these clumps is on the order of a few Myr, which is similar to their cloud crushing time. The fact that these cloudlets can survive in the wind for {\tEq{\approx}{5}} adds weight to the argument that radiative cooling can sustain cold cloud growth even under strong ablation driven by ram pressure from the hot phase (see also \citealt{Gronke2018}). This is in contrast to the no-cooling case where we find no cold gas in the outflow (see Section \ref{sec:phase diagrams}).

\begin{figure*}
    \centering
    \includegraphics[width=0.98\textwidth]{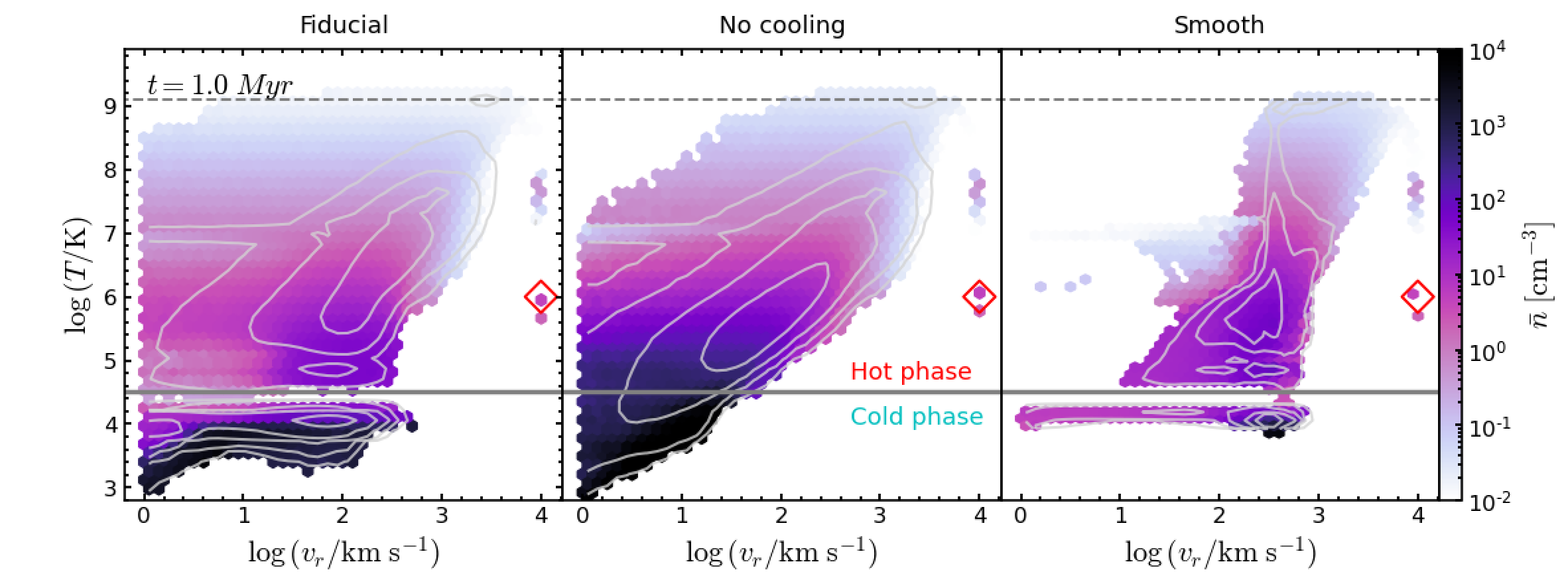}
    \caption{Phase diagram showing the gas temperature against the radial velocity. Bins are coloured by the mean number density in each bin, with the contours showing lines of constant mass, increasing logarithmically. The red diamond marks the initial wind injection, which quickly shocks to a post-shock temperature of {\TEq{=}{9.1}}, matching the analytic expectation (dashed line; \citealt{Costa2020}). The left panel shows the fiducial simulation, with medium clumps and radiative cooling; the middle panel shows the same initial clumps, but without cooling; and the right panel shows a smooth disc, with cooling. The horizontal grey line shows the cut between our `hot' and `cold' outflow definitions at \TEq{=}{4.5}. We can see that cooling is essential to forming a significant cold outflow, and that the velocity structure of both outflow phases is strongly dependent on whether the initial medium was clumpy or smooth.}
    \label{fig:phase}
\end{figure*}

\subsubsection{Comparison to a smooth disc} \label{sec:smooth_disc_qual}

Figure \ref{fig:smoothdisc} shows our simulations with an initially smooth disc for comparison with the clumpy setup. Many previous analytic and numerical studies consider a spherically-symmetric, homogeneous setup to study the propagation of AGN winds \citep[e.g.][]{King2003,Faucher-Giguere2012,Nims2015,Richings2018a}. Before disc break-out, our setup mirrors this regime. The temperature of the outflow is shown, with the time increasing in $0.1 \rm{\ Myr}$ increments clockwise from the top left. 
The wind sweeps up the gas in the disc into a thin shell of shocked ambient gas which, at \tEq{=}{0.7}, is travelling at a velocity of {\vrEq{\approx}{400}} with a temperature of {$T \approx 10^6 \rm{\ K}$} (top-left quadrant). By \tEq{=}{0.9} however, we see the emergence of a thin shell of cold (\TEq{\approx}{4}) gas, at a radius of {$R_{\rm{cool}}\approx 500 \rm{\ pc}$} (bottom-right). This is in accordance with the expected analytic result for an ambient density of $n_0\approx5\rm{\ cm^{-3}}$ (see Figure 1 in \citealt{Costa2020}). The high density of this shell results in a shorter cooling time, creating a multiphase outflow where the hot and cold components are expanding co-spatially with the same velocity. The sudden emergence of this cold shell introduces two outflow regimes: before post-shock cooling ($\tEq{\lesssim}{0.8}$) and after post-shock cooling ($\tEq{\gtrsim}{0.8}$). In the following analysis, we will often evaluate our results at \tEq{=}{0.5} and \tEq{\geq}{1} to represent both regimes. The outflow starts breaking out of the disc along the polar directions at \tEq{\approx}{0.5} which allows some hot gas to vent out at higher velocities and turbulently mix with the cold outflow (seen at the bottom of Figure \ref{fig:smoothdisc}). However, the overall outflow mass is still dominated by the spherical, expanding shell (see Section \ref{sec:phase diagrams}). This shell-like outflow is starkly different to the one seen in Figure \ref{fig:summary} where the cold gas is contained in small clouds entrained in a hot wind.

In this section, we have shown that a small-scale AGN wind model can produce kpc-scale, multiphase outflows. However, the morphology of this outflow depends strongly on the initial conditions: a smooth disc results in a thin shell of cold gas, formed by post-shock cooling, outflowing at a similar velocity as the shocked ambient medium. An initially clumpy setup results in high-velocity chimneys of hot gas, as the wind punches through regions of low density. The addition of cooling further results in the initial clumps fragmenting into small ($10-20\rm{\ pc}$) cloudlets which are then entrained in the hot outflow and can survive, or grow, on Myr timescales due to efficient cooling at their surface. These two setups, and their different cold gas formation channels, result in very different outflow energetics, which we quantify in Section \ref{sec:energetics} and discuss the implication of for observational estimates of outflow properties in Section \ref{sec:challenges_measure}.

\subsection{Multiphase gas energetics} \label{sec:energetics}
In this section, we analyse the multiphase structure of the outflows generated by our model (Section \ref{sec:phase diagrams}) before investigating how the outflow momentum and energy is distributed among the different gas phases (Section \ref{sec:cumulative}).

\subsubsection{Phase Diagrams} \label{sec:phase diagrams}

Figure \ref{fig:phase} shows the temperature of the gas as a function of radial velocity at {\tEq{=}{1}}. The pixel colour shows the mean number density, $\bar{n}\ \rm{[cm^{-3}]}$, within each pixel, and the logarithmic contours show the total mass of the outflow. We show these phase diagrams for three simulations (see Table \ref{table:runs}): `Medium clumps' (the fiducial model; left panel), `No cooling' (everything the same as fiducial, but without cooling; middle panel), and `Smooth' (homogeneous medium, with cooling; right panel). In our fiducial case (left panel) we see a clear two-phase outflow, with a cold, dense component at {\TEq{\lesssim}{4.5}} and a faster moving hot component centred on {\TEq{\approx}{5.5-7.5}}. This clear division in the phase structure motivates our later cut between the cold and hot outflow components at {\TEq{=}{4.5}} (horizontal grey line). Looking first at the hot component, we can see the injected wind at {\vAGNEq{=}{10^4}} and {\TEq{=}{6}} is quickly shock-heated to high temperatures ({\TEq{\approx}{9}}), which is in accordance with the expected post-shock temperature of the injected wind (see Equation 17 in \citealt{Costa2020}, shown here as a dashed line). The bulk of the hot component (highest mass contour) is moving at a radial velocity of {\vrEq{\approx}{500-1000}} and has a low density ($\bar{n}\approx 0.01-1 \rm{\ cm^{-3}}$), especially at high temperatures ({\TEq{\gtrsim}{7}}). Intriguingly at {\TEq{\approx}{5}}, we see a clear enhancement in density at a velocity of {\vrEq{\approx}{100}}, to around $ \bar{n} \approx 100 \rm{\ cm^{-3}}$, which is much more dense than any of the hot gas in the initial setup. The cold ({\TEq{\leq}{4.5}}) outflow moves more slowly than the hot phase, up to a radial velocity of {\vrEq{\approx}{400}}. The number density of this component is high, up to $\bar{n} =10^4\ \rm{cm^{-3}}$ which is similar to colder gas phases traced in observations (discussed further in Section \ref{sec:emission_lines}). The cold gas has a wide range of outflow velocities and is moving slower than the hot phase, as it is the densest gas, which is difficult for the wind to accelerate. 

The middle panel of Figure \ref{fig:phase} shows the phase structure for the same initial conditions as the fiducial simulation, but without cooling. It can be seen in this case that the AGN outflow does not contain a significant cold outflow phase -- there is no cold gas moving at any velocities higher than {\vrEq{\gtrsim}{40}}. This highlights that radiative cooling of the cold outflow is essential for its survival, as the AGN wind cannot simply push out the existing cold clumps without destroying them. This matches literature results which struggle to generate a cold outflow that can survive on Myr timescales without efficient radiative cooling \citep[e.g.,][]{Klein1994,Costa2015}.

The third panel of Figure \ref{fig:phase} shows the phase structure for a smooth disc. As mentioned in Section \ref{sec:smooth_disc_qual}, the formation and evolution of the multiphase outflow differs markedly from the clumpy case. The time shown (\tEq{=}{1}) is after the post-shock cooling regime, so we have both a cold (\TEq{<}{4.5}) and hot phase. There is a clear velocity peak in both phases at \vrEq{\approx}{400}. This differs from the clumpy case where both phases have a broader range of velocities, and the hot phase moves significantly faster than the entrained cold clouds. In the smooth case, we do see some hot gas moving at speeds \vrEq{\approx}{1000}, but this is due to the small amount of gas leaking out of the edge of the disc (bottom of Figure \ref{fig:smoothdisc}) and the fast small-scale wind in the centre. Nevertheless, the bulk of the outflowing mass remains moving at the characteristic velocity of \vrEq{\approx}{400}. The peak density of the outflow is in the cold shell, and is similar to the peak density in the clumpy case of $\bar{n} \approx 10^4 \ \rm{\ cm^{-3}}$.

\begin{figure}
    \centering
    \includegraphics[width=0.44\textwidth]{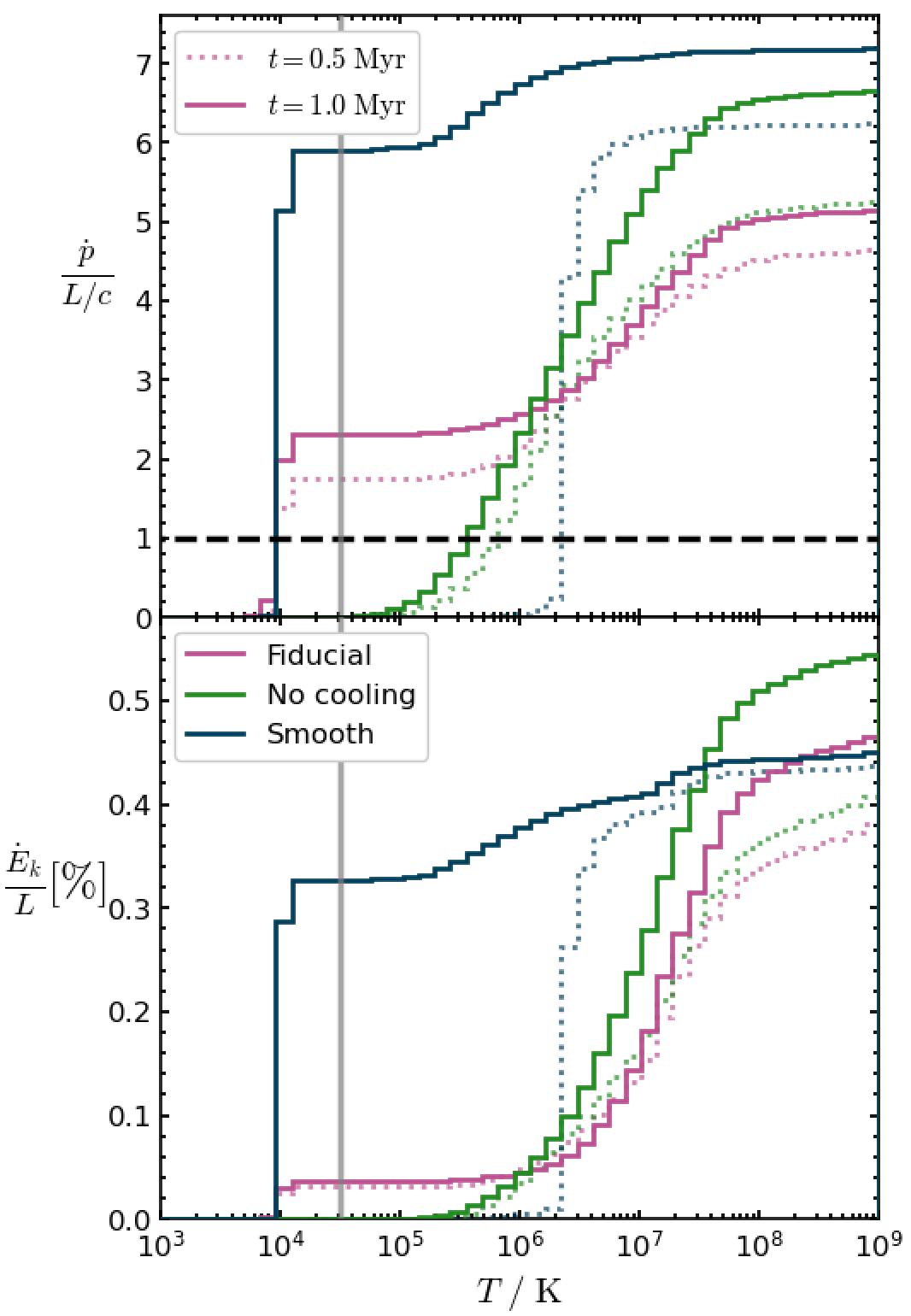}
    \caption{Cumulative momentum rate (top panel) and kinetic energy coupling efficiency (bottom panel) as a function of temperature. The dotted lines show the simulation at {\tEq{=}{0.5}} and the solid lines at {\tEq{=}{1}}; the colours represent initial conditions of medium clumps with radiative cooling (pink) and without cooling (green), and a smooth disc (blue). The vertical grey line shows our hot/cold phase split at {\TEq{=}{4.5}} and the horizontal dashed line shows $\dot{p}/(L/c)=1$ which is commonly used to differentiate the energy- and momentum-driven regimes. In the fiducial case, the hot outflow dominates the energy budget due to its higher velocity, whereas in a smooth disc, the cold outflow contains a significant fraction of the momentum and energy.}
    \label{fig:cumulative}
\end{figure}

\subsubsection{Energetics distribution} \label{sec:cumulative}

We now investigate the proportion of the radial (scalar) momentum and kinetic energy of the outflow contained in the different phases for the same simulations as in the previous section. Figure \ref{fig:cumulative} shows the cumulative momentum flux ({\pEq{}{}}; top panel) and kinetic coupling efficiency ({\Ek}; bottom panel) as a function of temperature. We show the global outflow quantities, calculated based on a velocity cut of {\vminEq{=}{10}} (Section \ref{sec:method_outflow}). The dotted lines show a time of {\tEq{=}{0.5}} and the solid lines show {\tEq{=}{1}} (shown to represent before and after the post-shock cooling in the smooth disc case). The vertical line shows our cold/hot phase split of {\TEq{=}{4.5}}. There is negligible difference in the momentum or kinetic energy fluxes carried in the range {\TEq{=}{4-5}}, demonstrating that our results are insensitive to the exact choice of temperature cut.

Results for our fiducial simulation (medium clumps and cooling) are shown in pink in Figure \ref{fig:cumulative}. The time evolution of this system is slight, with an increase in the momentum flux of $\approx \! 0.5$ in both the cold phase and hottest ({\TEq{\gtrsim}{7}}) gas, and an increase in the energy coupling of $0.1$ per cent in the hottest gas. We can see that, at {\tEq{=}{1}}, the cold phase (\TEq{<}{4.5}) has a momentum boost of {\pEq{\approx}{2.4}}, which is in the energy-driving regime ({\pEq{>}{1}}, horizontal dashed line). The total momentum boost rises to {\pEq{\approx}{5}} when we include the hot gas, showing that the momentum is split roughly evenly between the two phases. However, when we look at the kinetic coupling efficiency, we can see that the hot phase is dominant: the cold phase only has a coupling of {\EEq{=}{0.04}}, compared to the total of {\EEq{=}{0.4}}, i.e. the hot gas contains $90\%$ of the kinetic energy of the system. The reason for this difference in energy balance between the phases when compared to the momentum boost is the 5-10 times higher velocity of the hot outflow with respect to the entrained cold gas clouds.

For comparison with our fiducial simulation, the no-cooling case is shown in Figure \ref{fig:cumulative} in green. As seen in Figure \ref{fig:phase}, there is no cold outflow without cooling, so the energy and momentum are all contained in gas with {\TEq{\gtrsim}{5.5}}. Again, there is a mild time dependence between the two times shown, with the increase mostly occurring in the hottest gas. Comparing the no-cooling to the cooling simulation, we can see that the inclusion of radiative cooling reduces the momentum boost from {\pEq{=}{6.5}} to {\pEq{=}{5}}, representing a $23\%$ loss, and the kinetic energy coupling rate from {\EEq{=}{0.5}} to {\EEq{=}{0.4}}, representing a $20\%$ loss, at {\tEq{=}{1}}. This reduction in energy rate is both due to radiative losses and is also the result of the outflow having done different amounts of PdV work, due to the different structure of the outflows in the cooling/no cooling cases (see Figure \ref{fig:summary}). However, this reduction in momentum boost in the cooling simulation is not enough to change the outflow solution from energy- to the momentum-conserving, as {\pEq{>}{1}} still holds, although more efficient cooling (i.e., via metal- or molecular-lines) could reduce these values further. The exact reduction in momentum boost and energy rate when cooling is included may also be dependent on the ISM structure, with different sized clouds affecting how much ablation and mixing can occur between the wind and the ISM. The effect of initial clump size is discussed further in Section \ref{sec:kmin_vr}.

The dark blue line in Figure \ref{fig:cumulative} shows the resulting outflow from an initially smooth disc. As discussed in Section \ref{sec:phase diagrams} the cold gas formation mechanism in outflows propagating in a homogeneous medium is post-shock cooling, which results in a phase transition at the cooling time (in our case, at {$t_{\rm{cool}}\approx 0.8 \rm{\ Myr}$}, see Figure \ref{fig:smoothdisc}). Thus the time evolution of this system is much more significant than the clumpy case. At {\tEq{=}{0.5}} there is no cold outflow whereas at {\tEq{=}{1}} the cold outflow dominates both the momentum and energy coupling (containing {$85 \%$} of the momentum flux and {$75 \%$} of the kinetic luminosity). This is in stark contrast to the clumpy case where the kinetic energy in the cold gas is negligible ({$10\%$} of the total energy). This is because most mass in both the cold and hot outflow in the smooth case is confined in a thin shell  (Figure \ref{fig:smoothdisc}), expanding at {\vrEq{\approx}{400}} (Figure \ref{fig:phase}). However, in the clumpy case, the hot gas is able to vent much faster (up to {\vrEq{\approx}{1000}}) past the entrained cold clouds (Figure \ref{fig:summary}) which are generally travelling much slower (around {\vrEq{\approx}{10-100}}; Figure \ref{fig:phase}). This results in much less efficient transfer of momentum and energy from the hot AGN wind to the cold component of the outflow. The overall momentum of the smooth case is also higher, at {\pEq{=}{7}}, however, the total kinetic energy is similar to the cooling clumpy case at {\EEq{=}{0.44}}.

To summarise, in this section we have shown that cooling creates a two-phase outflow, but the share of the momentum and energy carried by each phase strongly depends on the initial conditions of the disc. The outflows are all in the energy-conserving regime ($\dot{p}/(L/c)>1$), however, in the clumpy case there is significant mixing between wind fluid and ambient gas. The associated cooling gives rise to a cold outflowing phase, but does not cause sufficient cooling losses to drive the outflow solution into the momentum-driven regime. Furthermore, the clumpy media confines the energy-driven bubble less efficiently, resulting in lower momentum fluxes than in a smooth disc.

\begin{figure}
    \centering
    \includegraphics[width=0.44\textwidth]{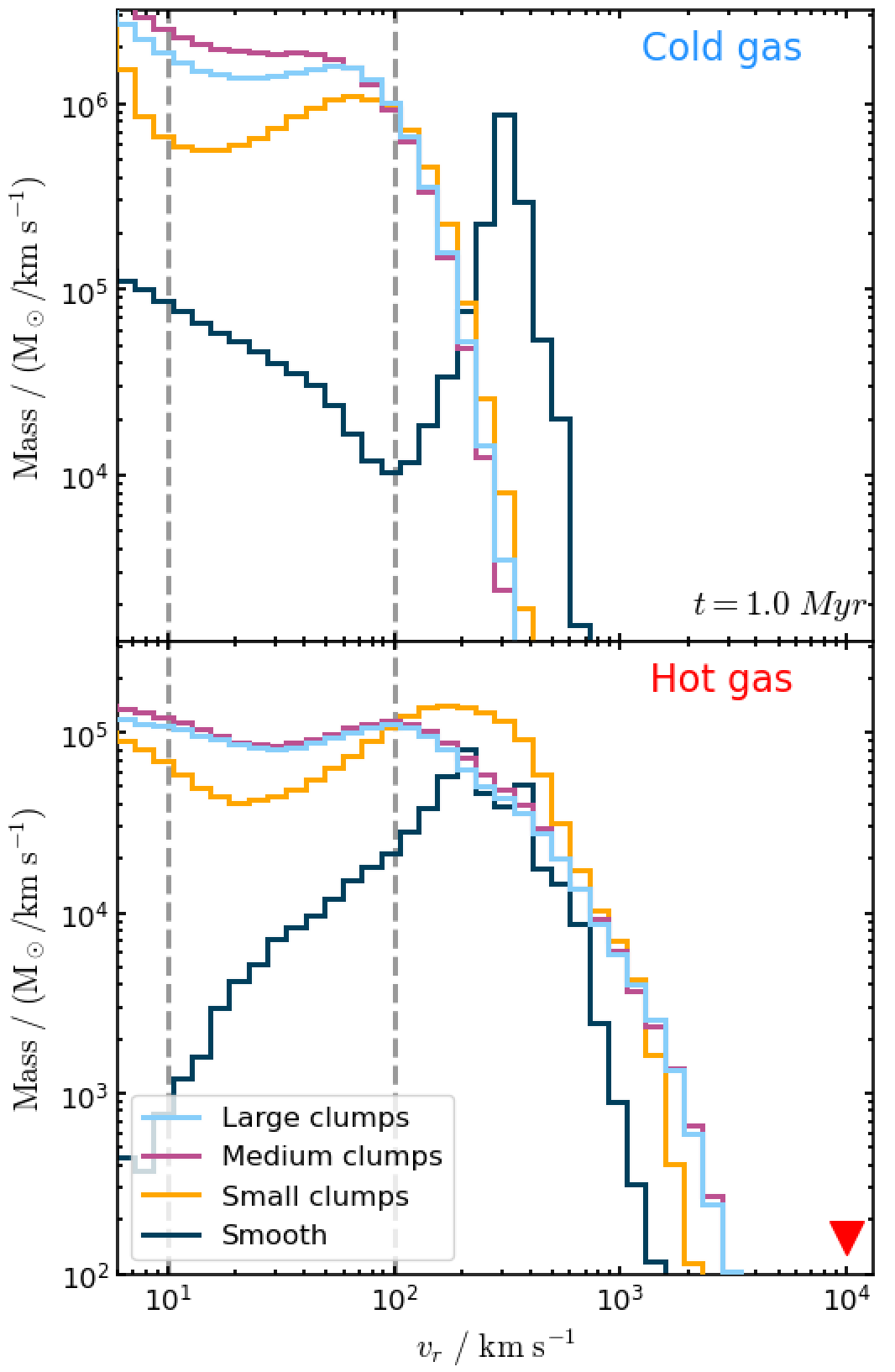}
    \caption{Histograms showing the outflowing mass at different radial velocities at {\tEq{=}{1}}. Each bin is normalised by its width to remove dependence on bin size. Gas is split by phase into cold gas (top panel) and hot gas (bottom panel). The colour of the lines represents the initial conditions of the simulation: large clumps (light blue); medium clumps (pink); small clumps (yellow) and smooth medium (dark blue). The red arrow marks the velocity of the injected AGN wind and the dashed grey vertical lines show the two velocity cuts used (\vmin). The smooth case shows a clear characteristic velocity peak at at {\vrEq{\approx}{400}} in both the cold and hot phases, whereas the clumpy setup shows a range of outflow velocities, and a clear difference in velocities between the phases.}
    \label{fig:vr_hist}
\end{figure}

\subsection{Parameter Variation} \label{sec:variation}

In Section \ref{sec:energetics}, we explored how the outflow produced from interactions between a clumpy ISM and an AGN wind differs from that in a smooth disc. We will now briefly analyse the sensitivity of our results to changes in the initial conditions of the simulations to assess the robustness of our conclusions. We investigate the impact of: the choice of distribution of clump sizes in the ISM initial conditions ({\LmaxEq{=}{40,170,330}}; see Figure \ref{fig:method}); a different choice for the density and thickness of the disc (and hence the total mass); and the speed of the wind at injection. Variations in AGN luminosity are explored later, in Section \ref{sec:fiore}. In Section \ref{sec:kmin_vr} we focus first on the initial ISM conditions and the impact this has on the radial velocity distribution of the outflow, and then in Section \ref{sec:sensitivity} we explore the change in mass outflow rate for the other simulation variations shown in Table \ref{table:runs}. We also test the numerical convergence of these global properties, finding that they are well-converged at our fiducial resolution of {\MtargetEq{=}{100}} (see Appendix \ref{ap_resolution}).

\subsubsection{Outflow variation due to initial clumpiness} \label{sec:kmin_vr}

Figure \ref{fig:vr_hist} shows the distribution of the outflowing mass as a function of radial velocity at $t=1\ \rm{Myr}$ for the three different initial clump sizes. The top panel shows the cold ({\TEq{<}{4.5}}) phase and the bottom shows the hot gas. Note that the y-scale on the bottom panel is smaller, demonstrating that the hot phase carries less mass in all but the highest velocity bins ({\vrEq{\gtrsim}{500}}). The red arrow marks the injection velocity of the AGN wind (not seen in the plot due to it's very low density).

We can clearly see, in both gas phases, that the differences between the clumpy cases (colourful lines) and the homogeneous medium (dark blue line) is greater than the variation between the initial clump sizes. As we are in the post-shock-cooling regime ({$t_{\rm{cool}}\geq 0.8\rm{\ Myr}$}; see Section \ref{sec:smooth_disc_qual}), the smooth case has produced a cold outflow which shows a narrow peak around {$\vrEq{=}{400}$}. The hot gas also peaks around a similar characteristic velocity, as it traces the shocked ambient gas shell. The shocked wind at small scales is also hot, but has very little mass. In contrast, the clumpy simulations show no such characteristic velocity, instead showing a range of speeds that gradually drop off after {\vrEq{\gtrsim}{100}}, for the cold phase, and {\vrEq{\gtrsim}{500}}, for the hot phase. This further demonstrates our earlier finding (Section \ref{sec:phase diagrams}) that there is a significant velocity differential between the phases when the initial medium is clumped, which is not seen in the homogeneous case.

We also find variations between the different initial clump size distributions (as also seen in \citealt{Wagner2012,Bieri2017}). Our medium ({\LmaxEq{=}{170}}; purple line) and large ({\LmaxEq{=}{330}}; blue) clumps show only small differences, despite the size of the average largest clumps being twice that of those in the medium case. The small clumps ({\LmaxEq{=}{40}}) show the greatest difference: there is a weak peak in velocity at {\vrEq{\approx}{80}} in the cold phase and {\vrEq{\approx}{200}} in the hot phase. There is also the lowest mass of slow-moving ({\vrEq{<}{100}}) cold gas compared to the other two clumpy simulations, with {$\sim \! 0.6\ \rm{dex}$} less outflowing mass above {\vrEq{=}{10}}. However, at radial velocities {\vrEq{>}{100}}, the small clumps case has the highest outflowing mass in all velocity bins. A similar picture is seen in the hot phase, with the small clumps having the lowest outflowing mass at {\vrEq{<}{100}}, but the most in the higher velocity gas ({\vrEq{=}{100-1000}}). However, in the fastest moving gas ({\vrEq{>}{1000}}), the small clumps case again drops to the lowest. An interesting point to note is that the lines for the small clumps (yellow) start trending towards the smooth case (dark blue) compared to the other two clump sizes, such as showing a more peaked distribution. This could suggest that many small clumps ({\LmaxEq{=}{40}}) start acting like a smooth medium in trapping the outflow and could contain a mix of post-shock-cooling shell and mixing layer cooling mechanisms. This restricts the fastest-moving hot gas, but is more effective at accelerating the cold cloudlets to velocities {\vrEq{>}{100}}. Furthermore, the resulting morphology of the cold clouds produced may still be dependent on the initial clumps from which they form; i.e. we may see smaller average cloudlets being formed from smaller initial clumps. These more detailed investigations are left to future work (Almeida et al, in prep.).

\begin{figure}
    \centering
    \includegraphics[width=0.4\textwidth]{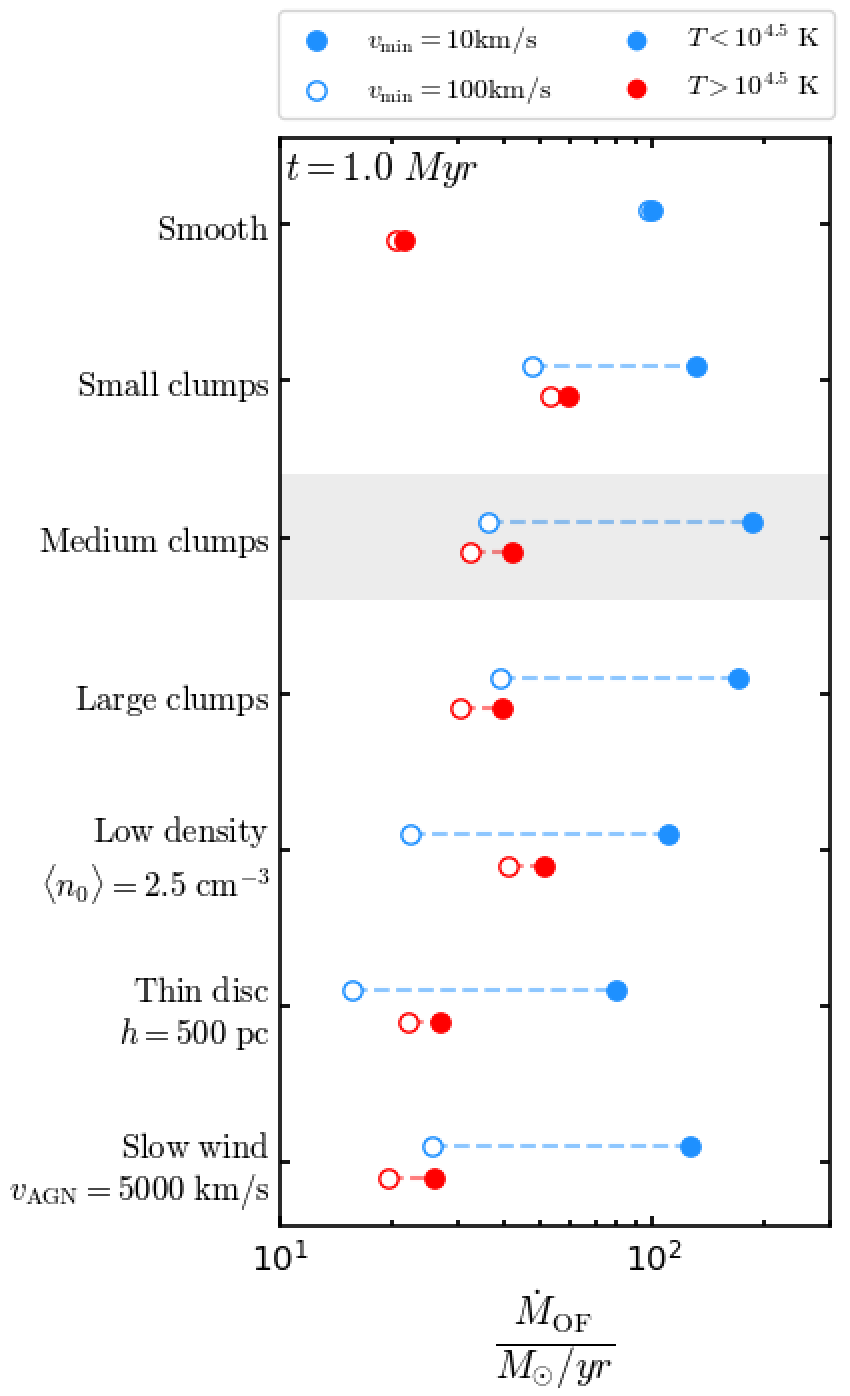}
    \caption{The mass outflow rates at {\tEq{=}{1}} for the range of simulation parameters explored, with our fiducial simulation shaded for reference. The cold phase ({\TEq{<}{4.5}}) is shown in blue and the hot phase ({\TEq{>}{4.5}}) in red. Two different outflow definitions are shown based on a radial velocity cut: {\vminEq{=}{10}} is shown in solid points and {\vminEq{=}{100}} are shown as empty circles. Despite the wide range in initial conditions shown, the outflow masses generally vary by less than a factor of two. However, the choice of {\vmin} has a much greater effect, with the cold phase outflow rates varying by around a factor of eight for all setups except the smooth disc.}
    \label{fig:param}
\end{figure}

\subsubsection{Sensitivity to disc setup} \label{sec:sensitivity}

Figure \ref{fig:param} shows the mass outflow rate at {\tEq{=}{1}} for a range of setup variations, as an example of the broad quantitative effect on outflow properties of our setup choices (this analysis was also performed for {\Ek} with similar results). In this Section, we restrict ourselves to only analysing simulations with {\LbolEq{=}{45}} and examine differences due to AGN luminosity to Section \ref{sec:fiore}. We use our two-phase cut with blue and red representing the cold and hot gas respectively. As discussed in the previous section, when the initial medium is clumpy, the outflow does not have a characteristic velocity, showing a range of velocities (Figure \ref{fig:vr_hist}). For this reason, we show two outflow definitions using differing radial velocity cuts: {\vminEq{=}{10}} as solid points and {\vminEq{=}{100}} as empty circles. The observational motivation and implications of these two cuts is discussed further in Section \ref{sec:outflow_v}. The following discussion will mostly focus on the {\vminEq{=}{10}} definition (solid points).

Our fiducial simulations (medium clumps) has a global mass outflow rate (following Section \ref{sec:method_outflow}) in the cold gas of $185 \rm{\ M_\odot / yr}$ and a hot mass outflow rate of $\approx 40 \rm{\ M_\odot / yr}$ (both at a velocity cut of {\vminEq{=}{10}}). This represents a mass outflow rate of almost a factor of two higher than the respective phases in the smooth case. This reiterates the point that, although the smooth case has greater momentum and energy fluxes (Figure \ref{fig:cumulative}), the clumpy setup creates a greater mass in the outflow, albeit moving at a slower velocity (Figure \ref{fig:vr_hist}). The large clump setup ({\LmaxEq{=}{330}}) has similar values to the fiducial case, with outflow rates lower by $\lesssim 10 \rm{\ M_\odot / yr}$. The small clumps case ({\LmaxEq{=}{40}}) has slightly elevated hot gas ($ 60 \rm{\ M_\odot / yr}$) but reduced cold gas ($130 \rm{\ M_\odot / yr}$) in its outflow, suggesting that smaller initial clouds lead to a higher trapping efficiency and/or an increase in mixing and cooling between the phases. Overall, varying the initial clumpiness of the disc changes the outflow rate by up to $ \lesssim 30\%$.

We also varied the initial mass of the galaxy, by decreasing both the initial mean density and the height of the disc by a factor of two compared to the fiducial case. The low density setup ($\langle n_0 \rangle = 2.5 \rm{\ cm^{-3}}$) has a slightly reduced mass outflow rate compared to the fiducial case ($110 \rm{\ M_\odot / yr}$) and shows a marginal ($10 \rm{\ M_\odot / yr}$) increase in the hot phase. However, lowering the initial density also increases the porosity of the disc due to more regions falling above the $T_{\rm{crit}}$ cut (see Section \ref{sec:ISM_setup}), which slightly complicates the comparison here (see \citealt{Wagner2012}). The thinner disc has just under half the outflow rate in the cold phase compared to the fiducial case, and the hot phase is reduced by $\approx 10 \rm{\ M_\odot / yr}$. This reduction is mostly in the polar direction as there is less gas for the wind to interact with before it breaks through the top and bottom of the disc. These results for the varying the disc height are consistent with the work of \cite{Wagner2013} when they moved from a (thin) disc to a (thick) bulge distribution for the gas clumps. Finally, a slower AGN wind ({\vAGNEq{=}{5000}}) results in a mass outflow rate which is lower by $\approx 30\%$, due to the injection energy also being reduced by half (see Equation \ref{eq:E_k_init}). Such difference in galaxy disc structure and ISM conditions would add some modest scatter to observationally-derived scaling relations between the AGN power and outflow properties.

In conclusion, we have shown that changes in the ISM structure and galaxy setup change the mass outflow rate by up to a factor of two. We performed the same analysis on the momentum and energy rates and found similar trends. In Section \ref{sec:fiore}, we place these differences in the context of global scaling relations to further investigate how great an impact they have on our final results. Furthermore, the phase and multiscale structure shown in Figures {\ref{fig:summary} \& \ref{fig:phase}} are qualitatively consistent across all the clumpy simulations, and in contrast with the smooth disc setup (Figure \ref{fig:smoothdisc}). However, a large factor in determining the quantitative outflow rates is the choice of {\vmin} cut due to the relative distribution between phases. An increase in this value from {\vminEq{=}{10}} to {\vminEq{=}{100}} can lead to a reduction of the measured cold mass outflow rate by a factor of eight, as seen by the empty circles, which we will explore in more detail in Section \ref{sec:outflow_v}.

\section{Results II: Implications for observed outflow properties} \label{sec:implications}

In the previous section, we explored how the interaction between an AGN wind and an initially clumpy medium creates a multiphase, multiscale outflow, which is qualitatively distinct from the spherical shell-like outflow generated in the homogeneous disc. In this section, we discuss the implications of our results for observational studies of AGN outflows. We review how the nature of these clumpy outflows poses challenges for measuring outflow properties such as outflow rates (Section \ref{sec:challenges_measure}) and also explore how this affects whether the outflow is inferred to be energy- or momentum-driven (Section~\ref{sec:challenges_driving}). Finally, in Section \ref{sec:fiore}, we investigate scaling relations between outflow properties and AGN luminosity and compare these to recent observational attempts.

\subsection{Challenges for measuring outflow properties} \label{sec:challenges_measure}

When assessing the potential impact of AGN on the ISM and host galaxy evolution, observers take multiple approaches \citep[see review in][]{Harrison2024}. One common approach is to calculate the time-averaged mass outflow rate, which, in its simplest form, is given by
\begin{equation}
    \dot{M}_{\rm{OF}} = B  \cdot \frac{M_{\rm{OF}}\cdot v_{\rm{OF}}}{R_{\rm{OF}}} \, ,
    \label{eq:S4_massoutflow}
\end{equation}
where, usually, $B=1$ or $B=3$, depending on the assumed geometry of the outflow \citep[see e.g.,][]{Gonzalez-Alfonso2017}. The momentum flux and kinetic luminosity\footnote{We note that, sometimes, an additional term is added to the kinetic luminosity, to account for turbulent gas motions.} of the outflows are then calculated with increasing powers of velocity, respectively as
\begin{equation}
    \dot{p} = \dot{M}_{\rm{OF}}\cdot v_{\rm{OF}} \, ,
    \label{eq:S4_momentum}
\end{equation}
and
\begin{equation}
    \dot{E}_{k} = \frac{1}{2}\cdot \dot{M}_{\rm{OF}}\cdot v_{\rm{OF}}^2 \, .
    \label{eq:S4_energy}
\end{equation}

These equations are then used to assess the potential impact of outflows, and underlying driving mechanisms, by normalising by the AGN luminosity to create the dimensionless quantities of momentum flux, $\dot{p}/(L/c)$, and outflow kinetic coupling efficiency, $\dot{E}_k/L$. These equations are commonly used to explore multiphase outflow properties across different observed galaxies populations, with the benefit that the energetics of the outflow can be measured using just the three quantities of $M_{\rm{OF}}$, $v_{\rm{OF}}$ and $R_{\rm{OF}}$ (as well as {\Lbol} for normalisation). However, it is important to assess their validity \citep[see derivations and discussion of challenges in e.g.,][]{Rupke2005b,Gonzalez-Alfonso2017,Harrison2018,Lutz2020,Veilleux2020,Kakkad2020}, and detailed observations and modelling are challenging some of their assumptions \citep[e.g.,][]{Crenshaw2000,Meena2023}. 
In this Section, we therefore explore using Equations \ref{eq:S4_massoutflow}, \ref{eq:S4_momentum}, \& \ref{eq:S4_energy} to compute outflow properties in line with these observational approaches (see Section~\ref{sec:challenges_driving}).

\subsubsection{Outflow velocity \& mass} \label{sec:outflow_v}

The bulk outflow velocity ($v_{\rm{OF}}$) is a critical measurement for inferring outflow rates (Equation \ref{eq:S4_massoutflow}). Observationally, outflow velocities are difficult to constrain \citep[see discussions in e.g.,][]{Veilleux2020,Harrison2018, Harrison2024} with issues including spectral resolution, projection effects and beam smearing \cite[e.g.,][]{Husemann2016,Wang2024}. One of the biggest challenges can be separating outflowing gas from non-outflowing gas. This is particularly problematic in spatially-unresolved observations, where a single emission line or absorption line profile is relied upon to determine an outflow velocity. There is a diversity of approaches in the literature for defining an outflow velocity from observations, and for deciding which gas is considered to be outflowing \citep[e.g.,][]{Carniani2015,Cresci2015, Harrison2018,Kang2018,Veilleux2020}. These include taking a minimum velocity threshold of the emission/absorption line wings to separate outflowing and non-outflowing material \citep[e.g.,][]{Kakkad2020,Lamperti2022}; taking the maximum velocity to represent the whole outflow to compensate for potential underestimation due to projection effects \cite[e.g.,][]{Fiore2017}; performing a multi-component fit, with the broad component tracing the outflowing gas \cite[see][for example explorations of different methods]{Lutz2020,HervellaSeoane2023,Gatto2024}; and constructing a dynamical model of the galaxy and assuming that the residuals are formed by  outflowing gas \citep[e.g.,][]{Rupke2017,RamosAlmeida2022,Girdhar2022}. However, all these methods, to greater or lesser extents, are still subject to systematic uncertainties and limitations on the minimum outflow velocity that can be determined above the systemic movement of the galaxy. This consequently has implications for estimating a total mass of outflowing material as masses are typically calculated by integrating the flux from a given line over a velocity range, and then converting this to a mass measurement.

In our simulations we thus explore using two radial velocity cuts to define which gas is outflowing: a `theoretical' value of {\vminEq{=}{10}} and a representative `observational limit' value of {\vminEq{=}{100}} (see Section \ref{sec:method_outflow}).
In Figure \ref{fig:param}, we show the mass outflow rate calculated both for {\vminEq{=}{10}} (solid circles) and {\vminEq{=}{100}} (hollow circles). 
We showed that if the outflow has a spherical shell-like morphology (smooth ISM case), the choice of {\vmin} has little impact on the resulting mass outflow rate. \textit{However, if the initial medium is clumpy, there is no single characteristic outflow velocity, with both phases showing a broad range in radial velocity distribution} (see Figure \ref{fig:vr_hist}) and thus no straightforward way to isolate AGN-driven outflowing motion. This results in the outflow mass being highly sensitive to {\vmin}, especially for the cold phase, as the bulk of this gas is travelling at lower velocities \citep[see also][]{Costa2018b}. Increasing {\vmin} from $10 \rm{\ km\ s^{-1}}$ to $100 \rm{\ km\ s^{-1}}$ results in around a factor of eight lower outflow rate for the cold gas, and around $10\%$ lower in the hot phase. This difference is much more significant than the variance in the measured outflow rate due to the initial condition parameters such as the clumpiness, disc height or AGN wind velocity (Figure \ref{fig:param}). For this reason, throughout this Section, we will consider the impact of using two different velocity cuts on outflow properties.

\subsubsection{Outflow phases}  \label{sec:emission_lines}
Another factor in measuring the mass of an outflow is which gas phase is detected. Although outflows have been detected in multiple phases \citep[e.g.,][]{Liu2013,Rupke2013,Carniani2015,Gonzalez-Alfonso2017,Girdhar2022,Girdhar2024}, observations are usually limited to tracing one phase, or only a few phases in a limited temperature range \citep[][]{King2015,Veilleux2020,Harrison2024}. As shown in Section \ref{sec:formation}, when the initial medium is clumpy, the cold phase ({\TEq{<}{4.5}}) carries the bulk of the mass (mass outflow rates a factor of 5 higher; Figure \ref{fig:param}), but the hot phase is much more energetic, with an order of magnitude higher kinetic energy coupling efficiency (Figure \ref{fig:cumulative}). These results are in agreement with observational studies, which typically find the colder outflow phases are lower velocity, less spatially extended, and have higher mass outflow rates \citep[][]{Vayner2021,Girdhar2022,Speranza2024}. However, studying the hottest outflow phases (i.e., {\TEq{\gtrsim}{6}}) is challenging due to the low density of the X-ray emitting gas, but there has been some success \citep[e.g.,][]{Greene2014,Veilleux2014,Lansbury2018}. \textit{The results presented in Section~\ref{sec:formation} clearly show that a multi-wavelength approach is crucial as otherwise a significant amount of mass/energy of the outflow will not be observed if only hot/cold phases are studied}. For the remainder of this section, we thus investigate the effect of separating our outflow into distinct phases through a simple temperature cut at {\TEq{=}{4.5}} (see Figure \ref{fig:phase}).

\subsubsection{Outflow radius and resolved outflow rates} \label{sec:radial}

\begin{figure}
    \centering
    \includegraphics[width=0.47\textwidth]{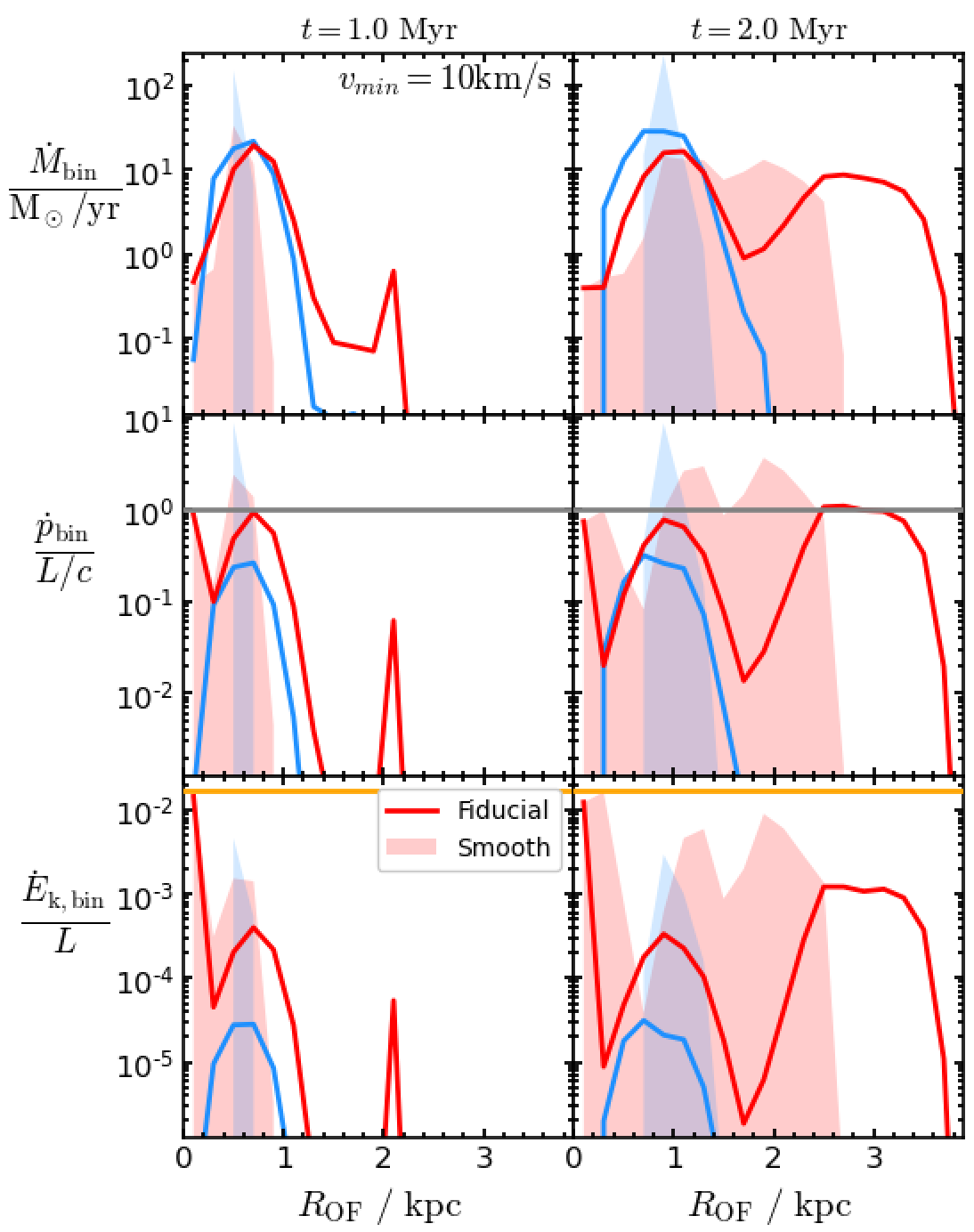}
    \caption{Radial evolution of the outflow, showing the localised mass outflow rate (top), momentum boosting (middle) and energy coupling efficiency (bottom) for the fiducial simulation (lines) and the smooth case (shaded). On the left is shown {\tEq{=}{1}} and on the right {\tEq{=}{2}}. The grey horizontal line in the middle panels shows the momentum-conserving value of $\dot{p}/(L/c)=1$ and the orange horizontal line in the bottom panels shows our injected energy rate ($\dot{E}_{\rm{k}}/L=1.7\%$). We can see that the cold phase (blue) dominates the mass at low radii, but the hot phase (red) dominates the energy, especially at large spatial scales.}
    \label{fig:radiusevo}
\end{figure}

The radial extent of detected AGN outflows varies from 10s pc - 10s kpc across multiple gas phases \cite[e.g., see][]{Veilleux2020,Harrison2024}. Measuring the exact value of $R_{\rm{OF}}$ can be challenging, especially in seeing-limited conditions, with a resolution of $1"$ corresponding to $2$ kpc at $z=0.1$ or $9$ kpc at $z=1$ \citep[e.g.,][]{Husemann2016,Tadhunter2018}, although spectroastrometric techniques can help improve the effective resolution \citep{Carniani2015,Lamperti2022}. If the outflow is unresolved, the beam size or fibre width is often taken as an upper limit on the radius. Some studies with good spatial resolution (order of $100 \rm{s\ pc}$) utilise long-slit spectroscopic or integral field unit observations to analyse the radial evolution of the outflow \citep[e.g.,][]{Crenshaw2000,Revalski2018,Revalski2021,Meena2023,Riffel2023}. In this Section, we emulate this approach by taking spherical bins of width $\Delta r = 200 \rm{\ pc}$ and calculating the total mass outflow, momentum and energy rate within each spatial bin \citep[following][]{Gonzalez-Alfonso2017} as

\begin{equation}
    \dot{M}_{\rm{bin}} = \frac{M_{\rm{bin}} \cdot v_{\rm{med}}}{\Delta r} \ ,
\end{equation}

where $M_{\rm{bin}}$ is the outflowing mass within each bin and $v_{\rm{med}}$ is the median velocity within the bin. Momentum and energy are calculated with further powers of $v_{\rm{med}}$. We take an outflow velocity cut of {\vminEq{=}{10}}, and note that because the calculation involves taking a median velocity, the results do not differ significantly from the higher {\vminEq{=}{100}} cut. The reduction in mass due to a higher {\vmin} cut is compensated by a higher median velocity. Our results are also insensitive to the bin width chosen. We stress that these \textit{radially resolved} values differ from the \textit{globally averaged} values used in the rest of the paper.

The radial profile of outflow properties is shown in Figure \ref{fig:radiusevo}. We show the mass outflow rate (top row), momentum flux (middle row) and kinetic energy coupling (bottom row) at outflow times of {\tEq{=}{1}} (left column) and {\tEq{=}{2}} (right column), with the colours showing the phase split as before (blue: cold, red: hot gas). The solid lines show the results from the fiducial (clumpy ISM) simulation and the shaded areas show the results for an initially smooth medium. 
In the clumpy case, we can see that the cold phase dominates the mass outflow rate, but only at small radii. The hot phase is moving faster and thus dominates the mass outflow rate at halo scales. The hot phase dominates the momentum flux and kinetic luminosity at all radii as its higher velocity more than compensates for its lower mass loading. 
This hot gas shows a two-humped structure, similar to that seen in the ionised phase of some observed outflows \citep{Revalski2018,Revalski2021}, which could point to two outflow structures: an equatorial outflow travelling at a slower speed and the uninhibited polar outflow. The outflow through the initially smooth disc (shaded areas; Figure \ref{fig:radiusevo}) is concentrated at a single radius, especially at {\tEq{=}{1}}, due to its shell-like morphology (see Section \ref{sec:smooth_disc_qual}; Figure \ref{fig:smoothdisc}). Interestingly, the kinetic energy coupling reached by the cold outflow in this smooth disc is two orders of magnitude higher than the maximum seen in clumpy case. This is likely due to its much more efficient confinement of the energetic hot gas (Section \ref{sec:smooth_disc_qual}).

A spatially-resolved approach shows that the two phases have very different radial extents, with the hot outflow reaching $r=3.5 \rm{\ kpc}$ by $\tEq{=}{2}$ and the cold outflow remaining within $r<1.5 \rm{\ kpc}$. As demonstrated in Figure \ref{fig:summary} and quantified in Figure \ref{fig:radiusevo}, \textit{when the initial medium is clumpy, there is no single characteristic outflow radius}, making it challenging to calculate single mass, momentum flux, or kinetic luminosity outflow rates, as is performed for observations where outflows are not spatially resolved.

\subsection{Challenges for interpreting momentum fluxes and kinetic luminosities} \label{sec:challenges_driving}

Alongside the challenges for measuring outflow properties, there are difficulties in using these quantities to interpret the resulting momentum boost and kinetic luminosities. Observational studies \citep[e.g.,][]{Cicone2014,Carniani2015,Fiore2017,Musiimenta2023} often use the momentum flux to assess whether the outflow is energy- ($\dot{p}/(L/c)>1$) or momentum-driven ($\dot{p}/(L/c)<1$). This is motivated from analytical models that predict that large-scale, energy-driven winds can have momentum fluxes as high as $\dot{p}/(L/c)\approx 20$ \citep{King2003,Zubovas2012, Faucher-Giguere2012}. Likewise, the kinetic energy coupling efficiency ($\dot{E}_k/L$) is often calculated in observations to assess how efficiently the AGN wind couples to the ISM \citep[e.g.,][]{Fiore2017}. The observed kinetic luminosity is sometimes compared to the input feedback efficiencies set in cosmological simulations, which generally are $\approx 5 \-- 20 \%$ \citep{Schaye2015,Weinberger2017}. Such comparisons are misleading, as these values represent the subgrid feedback efficiency assumed in the simulations -- which is often chosen for numerical, not physical reasons -- and cannot be straightforwardly translated into galaxy-wide outflow kinetic luminosities (see further discussion in \citealt{Harrison2018}). Although it may be possible for a perfectly energy-driven wind to achieve efficiencies of {\EEq{=}{5}} \citep{King2005}, in practice, the large-scale outflow may have an order of magnitude lower kinetic energy due to gravitational work and radiative cooling losses \citep{Veilleux2017,Richings2018b, Costa2020}.

In this Section, we explore the predictions of momentum boosts and kinetic coupling efficiencies from our simulations, and discuss the resulting challenges this might pose for observers in using these quantities for interpretation.

\subsubsection{Radial profiles of outflow properties} \label{sec:radial_evo}

In Figure \ref{fig:radiusevo}, we showed the spatially-resolved radial evolution of outflow  momentum and kinetic energy fluxes. We mark the commonly assumed boundary between the energy- and momentum-driven regime ($\dot{p}/(L/c)=1$) as a grey line in the second row of panels. In the clumpy case (solid lines), the cold outflow would always be seen as momentum-conserving, despite having a global value that places it in marginally energy-driven regime ($\dot{p}/(L/c)=2.5$; Figure \ref{fig:cumulative}). The hot outflow has consistently higher values (due to its higher velocity; Figure \ref{fig:vr_hist}) but would also still be interpreted as momentum-driven.
In the smooth case, however, the cold phase can be seen as energy-driven at $R_{\rm{OF}}=600\rm{\ pc}$ at {\tEq{=}{1}} and $R_{\rm{OF}}=1\rm{\ kpc}$ at {\tEq{=}{2}}. This is due to all the cold gas being constrained to a thin shell travelling at {\vrEq{\approx}{400}}. At {\tEq{=}{1}}, the hot phase is still mostly constrained to this shell so we also see an energy-driven outflow at $R_{\rm{OF}}=600\rm{\ pc}$, but by $\tEq{=}{2}$, enough has broken out of the disc to extend out to $R_{\rm{OF}}=2.2\rm{\ kpc}$.

The bottom row of Figure \ref{fig:radiusevo} shows the radial evolution of the kinetic energy coupling efficiency. The orange line in the bottom row shows the energy flux of the small-scale wind {\EEq{=}{1.7}}. The hot outflow completely dominates over the cold at all radii, with the cold outflow only having a peak value of {$\EEq{\approx}{0.003}$}. The hot outflow varies significantly with radius, but has a peak value of $\EEq{\approx}{0.1}$ in the outer shock front ($R_{\rm{OF}} \approx 3 \rm{\ kpc}$). The smooth case again has higher peak values, with the cold phase peaking at $\EEq{\approx}{0.4}$ and the hot at almost 1\% at $R_{\rm{OF}} \approx 2 \rm{\ kpc}$ at {\tEq{=}{2}}. Furthermore, there is a large variation in radius with these quantities, for example, the energy and momentum fluxes of the hot outflow drop by almost two orders of magnitude from $R_{\rm{OF}}=1 \rm{\ kpc}$ to $R_{\rm{OF}}=1.8 \rm{\ kpc}$ before rising by the same amount or more again at $R_{\rm{OF}}=2.4 \rm{\ kpc}$, once the outflow accelerates into the halo.

In conclusion, the location at which the outflow is measured can have a large effect on how the momentum fluxes and kinetic luminosities are interpreted. When studies spatially resolve the outflow, as we have emulated in Figure \ref{fig:radiusevo}, even if the global outflow is energy-driven, the wide spread in radii and velocities can result in no one radial bin implying an energy-driven solution based on measured momentum fluxes. For studies that do not spatially resolve the outflow and instead take a global approach, it is important to note that the two phases dominate at different spatial scales. In addition, a significant amount of energy and momentum may be missed when observations are unable to measure the hottest gas phases at the largest scales.

\subsubsection{Time evolution of global quantities} \label{sec:timeevo}

\begin{figure}
    \centering
    \includegraphics[width=0.43\textwidth]{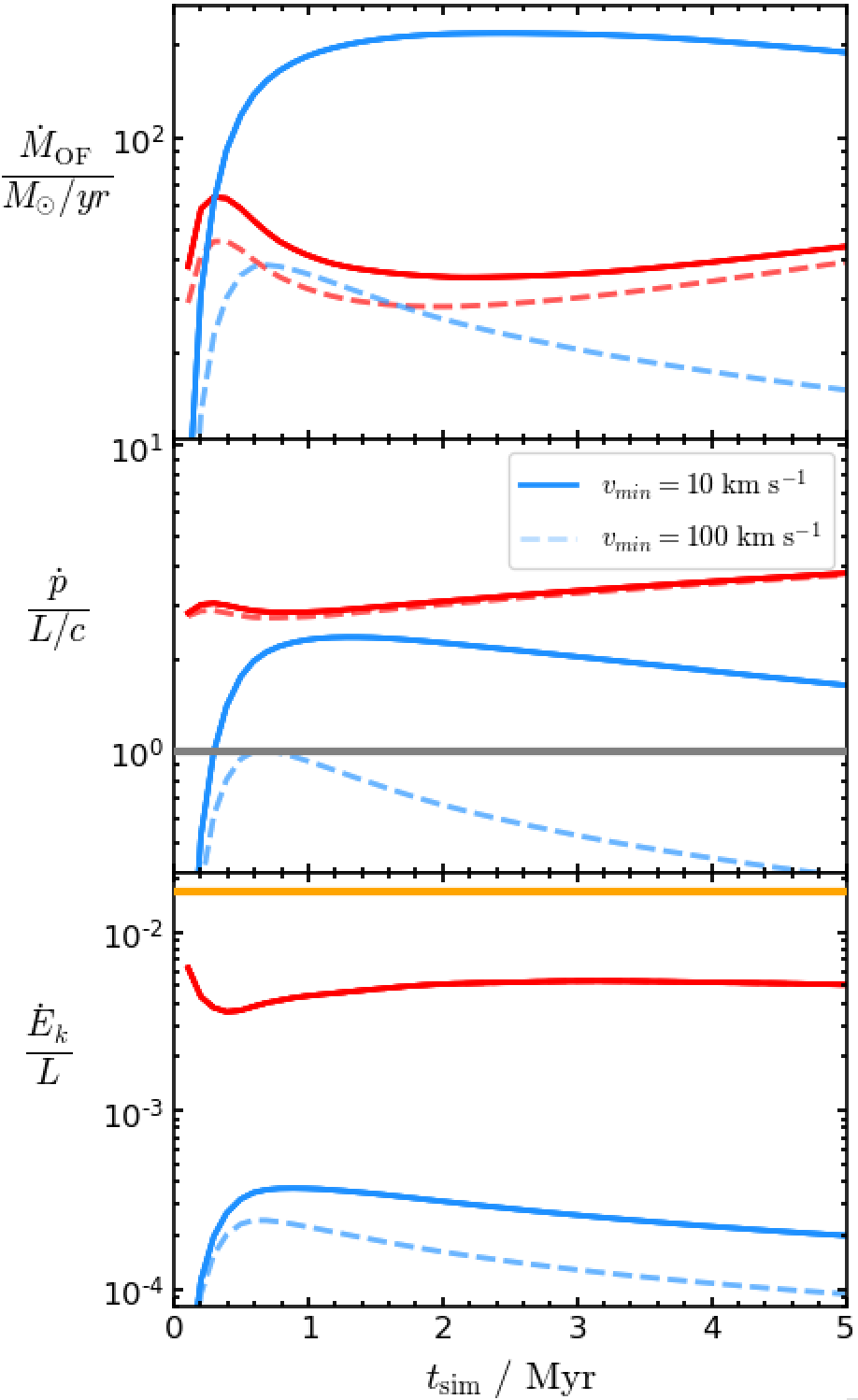}
    \caption{Time evolution of outflow in our fiducial simulation showing the mass outflow rate (top), momentum boosting (middle) and energy coupling efficiency (bottom). Solid lines show {\vminEq{=}{10}} and dashed lines show {\vminEq{=}{100}}. The cold phase is in blue and the hot phase is in red. The grey horizontal line (middle panel) shows the momentum-conserving value of $\dot{p}/(L/c)=1$ and the orange horizontal line (bottom panel) shows the injected energy rate ($\dot{E}_{\rm{k}}/L=1.7\%$). The cold phase dominates the mass outflow rate and the hot phase dominates the momentum flux and kinetic luminosity. However, the measured properties of cold phase is particularly sensitive to the minimum radial velocity cut ({\vmin}).}
    \label{fig:timeevo}
\end{figure}

We also investigated how the time at which the outflow is measured may affect how the global, time-averaged energetics are interpreted. 
Figure \ref{fig:timeevo} shows the global time evolution of the mass outflow rate (top panel), momentum boost (middle panel) and kinetic energy coupling efficiency (bottom panel), with the cold/hot phases in blue/red. 
The solid lines show results for outflow velocity cut of {\vminEq{=}{10}} (our `theoretical value') and the dashed lines show {\vminEq{=}{100}} (representative of an `observational' limit, see Section \ref{sec:outflow_v}). We can see that, using {\vminEq{=}{10}}, the cold outflow dominates the mass outflow rate, rising rapidly to a peak of \MofEq{\approx}{200} by {\tEq{=}{1.5}}, before flattening. The hot outflow shows a peak of \MofEq{\approx}{65} at {\tEq{=}{0.4}}, before settling to \MofEq{\approx}{40} by {\tEq{=}{1.5}}. However, using the higher outflow velocity cut of {\vminEq{=}{100}} drastically reduces the inferred cold gas outflow rate, which now peaks at {\MofEq{=}{40}} and shows a steeper tail-off, dropping by a factor of 2 in the $4 \rm{\ Myr}$ after the peak. As previously discussed (Section \ref{sec:outflow_v}), a higher outflow velocity cut has less impact on the hot phase, resulting in only a $\approx 15\%$ decrease across most of the simulation time.

A similar picture emerges when considering the momentum flux (middle panel of Figure \ref{fig:timeevo}). The momentum flux in the cold outflow again rapidly rises to a peak at \pEq{\approx}{2.5} by \tEq{=}{1} for {\vminEq{=}{10}}. However, when using the cut of {\vminEq{=}{100}}, the peak value reduces to $\dot{p}/(L/c) \approx 1$, before dropping further by 60\% by {\tEq{=}{5}}. We can see that both the outflow time and the minimum velocity cut used affect whether a momentum boost of \pEq{>}{1} or \pEq{<}{1} is measured. If only the cold phase is considered, the outflow may be incorrectly categorised as `momentum-driven' despite the overall outflow solution being energy-driven.


Looking at the third row of panels, the hot outflow dominates the energy of the outflow with a steady value of {\EEq{\approx}{0.5}}. However, the kinetic luminosity in the cold phase is much lower, peaking at \EEq{=}{0.04} and showing a strong time evolution, especially with the higher minimum velocity limit, dropping to \EEq{=}{0.01} at {\tEq{=}{5}}. Compared to the kinetic luminosity of the small-scale wind ({\EEq{=}{1.7}}; Equation \ref{eq:E_k_init}), we can see that the total kinetic luminosity of the large-scale outflow is lower by a factor of three. If only the cold phase were observed, this difference amounts to two orders of magnitude.

In this section we have shown that \textit{where}, \textit{when} and \textit{in which phase} the outflow is being measured can have a large impact on the inferred mass outflow rates, momentum boosts and kinetic coupling efficiencies. These challenges are exacerbated in the case of an inhomogeneous medium where there is a wider range of outflow velocities than in the smooth case. If only a single phase is measured, this can lead to the outflow being mischaracterised as momentum-driven despite the overall energy-driven nature of the outflow.


\subsubsection{Energy \& momentum conservation}

\begin{figure}
    \centering
    \includegraphics[width=0.48\textwidth]{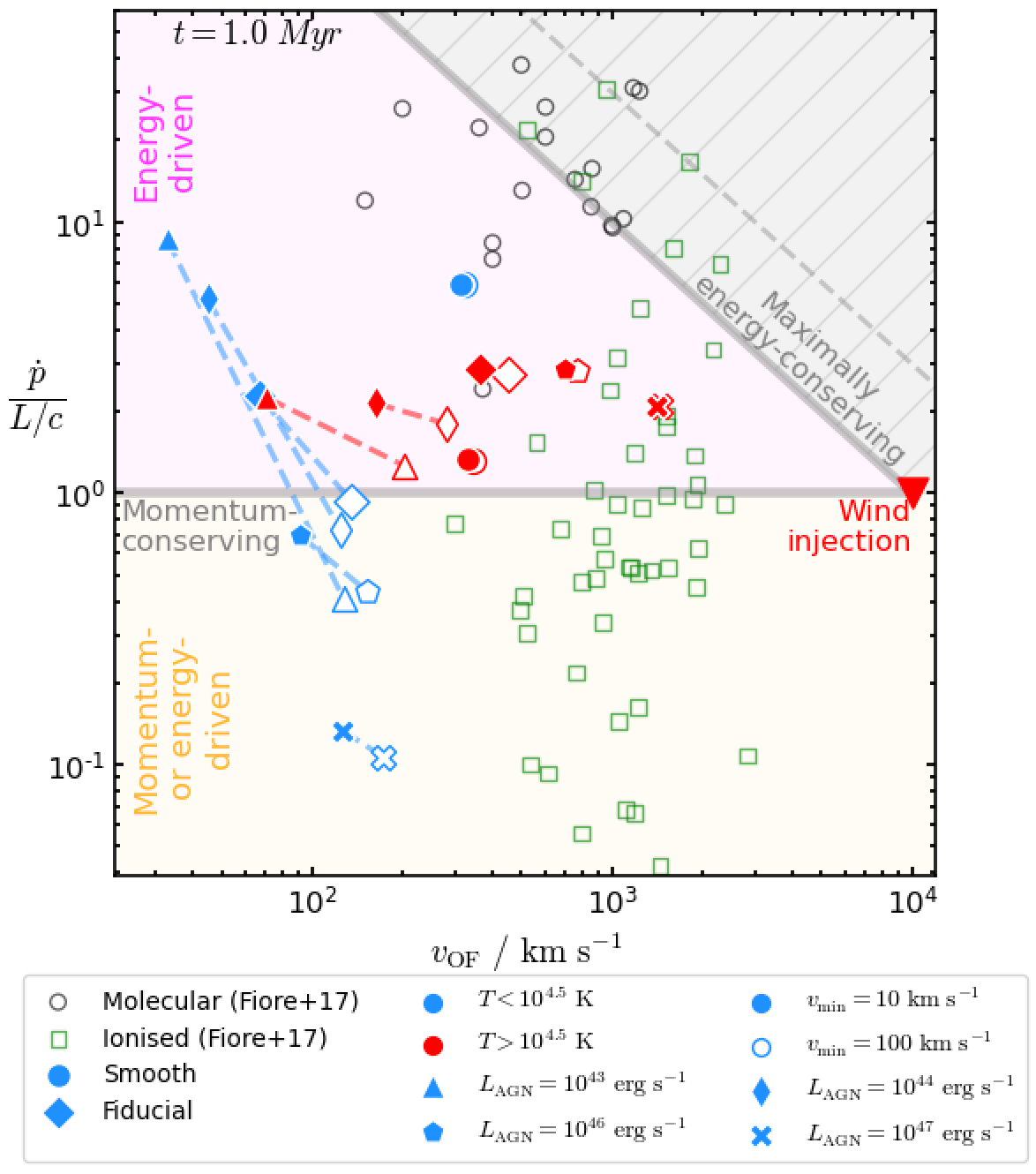}
        \caption{Outflow momentum flux against velocity for our simulation suite at {\tEq{=}{1}} for {\vminEq{=}{10}} (solid points) and {\vminEq{=}{100}} (hollow points). Blue/red represent the cold/hot phases and the shape of the point shows different AGN luminosities. Also shown are analytic expectations for energy- and momentum conservation (grey lines) based on our input velocity of {\vAGNEq{=}{10^4}}, with the dotted line showing the energy-conserving case for {\vAGNEq{=}{3\times10^4}}. These define the energy-driven (pink) and the momentum- or-energy driven (orange) regimes. The black circles and green squares show an observational compilation from \protect\cite{Fiore2017} for comparison, representing molecular and ionised gas outflows, respectively (although we caution against a direct comparison with our hot/cold phase split). We can see that, despite the outflow being energy-driven overall, measurements in just the cold phase can result in momentum fluxes \pEq{<}{1}.}
    \label{fig:mom_vel}
\end{figure}

Figure \ref{fig:mom_vel} shows momentum flux versus outflow velocity; a plane commonly used by observers to infer whether the measured outflow is momentum- or energy-conserving \citep[e.g.,][]{Bischetti2019,Marasco2020,Longinotti2023,Bischetti2024}. 
The grey lines represent analytic expectations, with the horizontal line showing the momentum-conserving case of $\dot{p}/(L/c)=1$. The diagonal lines shows a maximally energy-conserving outflow ($\dot{p}/(L/c) = v_{\rm{OF}}/v_{\rm{AGN}}$), representing the maximum momentum flux the large-scale outflow would have if it converted all the energy from the small-scale AGN wind. The solid diagonal shows the expectation for our fiducial AGN wind velocity of {\vAGNEq{=}{10^4}} and the dashed diagonal shows how this changes for a higher small-scale wind velocity of {\vAGNEq{=}{3\times10^4}}. We show the momentum flux of our small-scale wind as the red triangle at \vAGNEq{=}{10^4}. The other blue/red points show the momentum flux for the cold/hot gas at {\tEq{=}{1}}, calculated globally using Equation \ref{eq:S2_mom}. The outflow velocity is computed as a mass-weighted mean of the radial velocity for gas above a cut of {\vminEq{=}{10}} (solid points) and {\vminEq{=}{100}} (hollow points). 

For the fiducial simulation, the momentum fluxes are in the range \pEq{\approx}{2-3} for both the hot and cold phases, dropping to \pEq{\approx}{1} in the cold case for the \vminEq{=}{100} cut. As shown in Figure \ref{fig:vr_hist}, the hot phase has a higher velocity than the cold phase, with a mass-weighted mean velocity a factor of $4-5$ higher. We show results for a range of AGN luminosities, all with the fiducial disc parameters (Table \ref{table:runs}). 
The highest value for \pEq{}{} is measured in the lowest luminosity ({\LbolEq{=}{43}}) simulation, with \pEq{\approx}{9} in the cold phase. However, the velocities at these faint luminosities are low, resulting in a much greater difference when we vary {\vmin}. Even for the fiducial luminosity ({\LbolEq{=}{45}}), increasing the velocity cut to {\vminEq{=}{100}} results in the cold outflow having \pEq{\lesssim}{1}, which would result in it being misinterpreted as momentum-driven. Additionally, the circular points show the results for the smooth case. 
Here, both phases have approximately the same velocity, and the cold phase dominates the momentum flux with \pEq{\approx}{6}. We note that none of our simulated outflows reach the analytic expectation for maximal energy conservation (diagonal grey line). 
The closest point is the cold phase in the smooth outflow and this is still a factor of 5 times lower than the maximal value. 
Even if the global outflow produced by our model is energy-driven \citep{King2003,Costa2020}, the global momentum flux of the cold component can be low (\pEq{\lesssim}{1}), as it couples weakly to the clumpy ISM. 

In Figure \ref{fig:mom_vel} we show an observational compilation from \cite{Fiore2017} for comparison, with the black circles and green squares representing molecular and ionised outflows, respectively. For clarity, we do not show error bars on the observational points, but these are generally large, spanning around an order of magnitude. We are not attempting a one-to-one correspondence with the observations, so caution against a direct comparison. In particular, we stress that our hot/cold temperature split does not map directly to observed molecular/ionised phases, but should be considered as just a generalised trend for gas at different temperatures. Additionally, in this sample, \cite{Fiore2017} use the maximum outflow velocity to compensate for orientation effects, which differs from our mass-weighted mean approach. However, we can observe some general trends with gas phase. For example, we can see that the ionised outflows span a large range of \pEq{}{}, but the molecular outflows all have higher momentum fluxes, at or near the energy-conserving relation. This is in contrast to our simulations where we find that the inhomogeneous structure of the ISM results in weak coupling to the cold gas phase, resulting in modest momentum fluxes and lower outflow velocities. It is surprising therefore that these observed molecular outflows have such high momentum fluxes. This could be due to the observational assumption that all the gas is moving at the same velocity in a thin shell, which is in tension with our results which find entrained cold clumps outflowing at a range of radii. We note that even in an inhomogeneous ISM, there may be local conditions where the ambient medium behaves homogeneously, causing the outflow to propagate in a shell; for example if the AGN is embedded within a large cloud (see discussion in \citealt{Bieri2017}), or if the cold phase is arranged in a `mist' \citep{Gronke2020b} with a high covering factor that efficiently traps the wind (equivalent to an even smaller $\lambda_{\rm{max}}$ than in our simulations). However, overall, our results suggest that the cold outflowing phase does not readily show a shell-like morphology for the parameter space explored in this study.

\begin{figure*}
    \centering
    \includegraphics[width=0.94\textwidth]{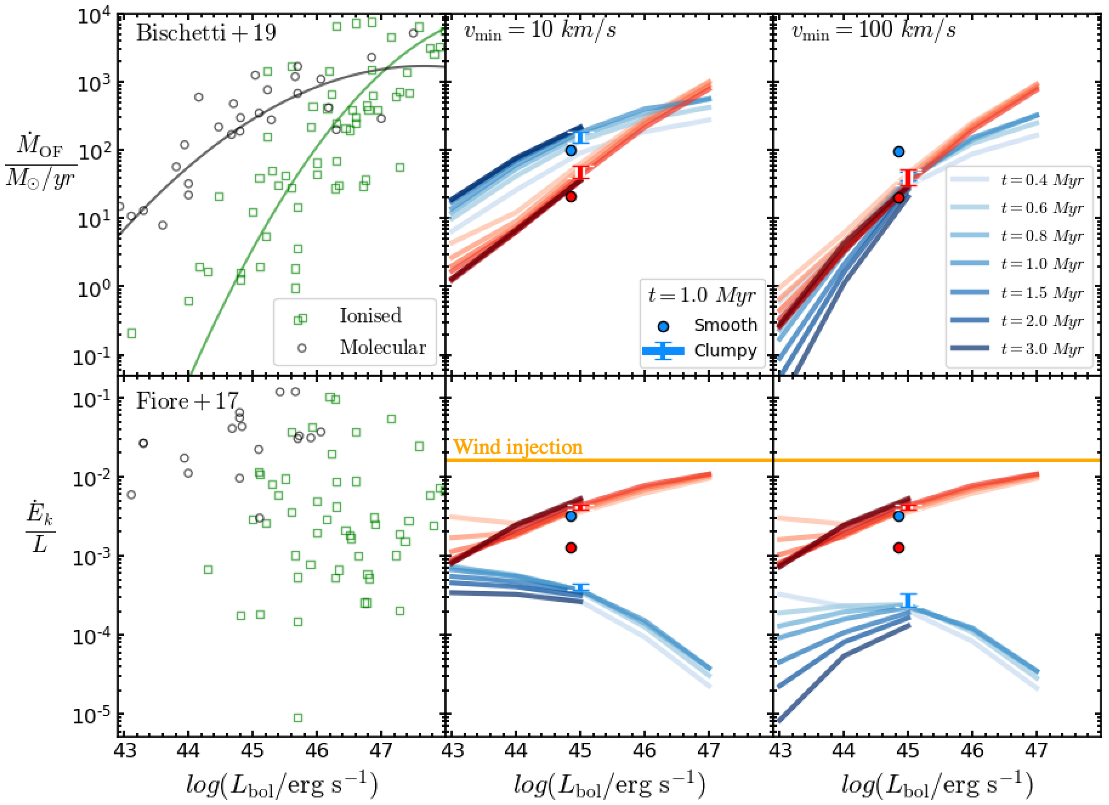}
        \caption{Scaling relations between the AGN luminosity, mass outflow rate and kinetic energy coupling efficiency. The blue/red lines show the cold/hot outflow in the fiducial simulation with darker lines showing later times. We note that the higher luminosity simulations ({\LbolEq{\geq}{46}}) were only evolved to \tEq{=}{1} due to computational and boxsize constraints). The errorbars show the minimum and maximum values across all the clumpy simulations at \tEq{=}{1} to show the expected scatter in properties due to the intrinsic ISM structure of the galaxy. The circular points show the result from the smooth case. The orange line (bottom panels) shows the small-scale injected energy rate ($\dot{E}_{\rm{k}}/L=1.7\%$). On the left, we show an compiled observational sample from \protect\cite{Fiore2017,Bischetti2019} for comparison, although we note that our hot/cold temperature split should not be directly compared to observed ionised/molecular phases. Our simulations predict a positive scaling relation between the mass outflow rate and AGN luminosity.}
    \label{fig:Fiore_ME}
\end{figure*}

\subsection{Outflow scaling relations with AGN luminosity} \label{sec:fiore}

One of the ways in which observational studies assess the role of AGN in launching outflows and the potential impact they have on galaxy evolution is to construct scaling relations between outflow, AGN and galaxy properties \citep{Fiore2017,Gonzalez-Alfonso2017,Leung2019,Bischetti2019,Lamperti2022,Musiimenta2023}. 
Some simulations have used these observational constraints as an input for their AGN models \citep{Rennehan2023}. 
As we have discussed, there are many difficulties and uncertainties when measuring these quantities related to characterising outflow velocities and masses (Section \ref{sec:outflow_v}), radii (Section \ref{sec:radial}), contributions from different phases (Section \ref{sec:emission_lines}), and variations in time (Section \ref{sec:timeevo}) and location (Section \ref{sec:radial_evo}). 
The fact that AGN vary faster than the outflow properties \citep{Zubovas2020}, and various target selection effects \citep[see discussion in][]{RamosAlmeida2022,Harrison2024} both pose additional challenges. 
There is therefore still debate about whether such scaling relations exist, with some studies finding them \citep{Fiore2017,Musiimenta2023} and others not finding tight correlations \citep{Davies2020,Lamperti2022}. 

\subsubsection{Simulations predict scaling relations}

Using the results of our fiducial simulations performed over the luminosity range {\LbolEq{=}{43-47}}, we calculate global mass outflow rates and kinetic coupling efficiencies (following Section \ref{sec:method_outflow}). We also explore the impact of variations in the initial disc clumpiness and explore results for a smooth medium at a single luminosity of {\LbolEq{=}{45}}. 
We present these results in Figure \ref{fig:Fiore_ME}, with the top row showing the mass outflow rate and the bottom showing the kinetic coupling efficiency as a function of AGN luminosity. The columns show our different outflow definitions: the middle shows the low velocity cut ({\vminEq{=}{10}}) and the right shows the higher velocity ({\vminEq{=}{100}}) outflow definitions. 
The solid lines show the results from our fiducial simulation, with the blue/red colour showing the cold/hot temperature cut and the gradient of the line showing the time evolution, with later times shown darker. 

Looking at the middle column in the top row ({\vminEq{=}{10}}), we can see that both gas phases have higher mass outflow rates with increasing {\Lbol} \citep[see also][]{Costa2020}. The hot phase shows the steepest correlation with the mass outflow rate increasing by 3 dex over the 4 dex luminosity range. The cold outflow rate also increases strongly at low {\Lbol}, but then begins to taper off, reaching a more pronounced turnover point at {\LbolEq{>}{46}}, which allows the hot phase to exceed it in outflow rate at the highest luminosities. This turnover possibly demonstrates a critical luminosity above which the AGN is too powerful to allow the cold clumps to survive in the wind. This result echoes \citet{Zubovas2017}, who identified a critical AGN luminosity of {$L_{\rm{AGN}}\approx 5 \times 10^{46} \rm{\ erg\ s^{-1}}$} at which AGN-induced cloud fragmentation is maximally effective and above which the wind ejects gas too efficiently for the gas to cool and fragment. 
There is some subtle time evolution, with the mass outflow rate in the hot phase decreasing with time and the cold phase increasing over the period \tEq{\approx}{0.5-3} (see Figure \ref{fig:timeevo}). However, this time evolution is subdominant compared to the overall positive trend with {\Lbol}.

For the high velocity cut {\vminEq{=}{100}}, the same qualitative trends remain, with both phases showing a positive correlation with an even steeper gradient. 
Raising the velocity cut has an even greater effect on the low-luminosity ({\LbolEq{=}{43-44}}) systems, with the cold mass outflow rate dropping by two orders of magnitude.
However, at the brighter end ({\LbolEq{>}{45}), {\vmin} plays less of a role and we still see the cold phase turnover at the same luminosity of {\LbolEq{=}{46}}. 

In Figure \ref{fig:Fiore_ME} we also show the results for an initially smooth medium as circular points. As shown in Figure \ref{fig:param}, a smooth medium results in a factor $\approx 2$ lower mass outflow rate. However, because the characteristic velocity of the resulting shell is higher than both choices of {\vmin}, the smooth case is unaffected by raising the minimum velocity cut (unlike the clumpy case).
It thus results in a cold outflow rate that is a factor of 3 larger than the clumpy case for {\vminEq{=}{100}}. To evaluate the contribution of initial ISM clumpiness on the resulting outflow properties, we show the contributions to the scatter on these scaling relations as the blue and red errorbars on the plot. These represent the maximum/minimum mass outflow rate of any of the simulations with {\LmaxEq{=}{40-330}} at {\tEq{=}{1}}. We can see that the change in mass outflow rate due to the initial clumpiness is smaller than the changes caused due to time evolution.

In the bottom row of Figure \ref{fig:Fiore_ME}, the orange line shows the kinetic luminosity of the injected small-scale wind ($\dot{E}_{\rm{k}}/L = 1.7\% $).
The hot phase (red) shows a slight positive correlation with {\Lbol}, increasing by 1 dex across the full AGN luminosity range probed by our simulations. 
The cold phase (blue), however, shows a negative correlation, with the trend accelerating downwards at higher luminosities {\LbolEq{>}{45}}. 
When taking the higher radial velocity cut (\vminEq{=}{100}), the kinetic coupling efficiency in lower {\Lbol} systems shows a strong decline with time. 
This time variation is most prominent at fainter AGN luminosities, suggesting that the intrinsic scatter in observed scaling relations should be highest for \LbolEq{\lesssim}{45}.
Finally, we note that the difference between the clumpy and smooth cases on the inferred outflow kinetic coupling efficiencies now becomes even more noticeable, with the cold phase kinetic coupling typically more than an order of magnitude higher in the homogeneous medium compared to the clumpy simulations.

\subsubsection{Observational comparison}

In the left panel of Figure \ref{fig:Fiore_ME}, we show observational compilations from \cite{Fiore2017} and \cite{Bischetti2019} split into measurements for molecular (black circles) and ionised phases (green squares). These have primarily been estimated based on CO and [\ion{O}{III}] line emission, respectively. 
The molecular phase dominates the mass outflow rate at lower luminosities, but flattens off around {\LbolEq{=}{46}}. The ionised phase increases consistently, becoming dominant in the highest-{\Lbol} systems. The observed kinetic luminosities do not show any obvious trend with {\Lbol}. The molecular phase has high ($>1\%$) coupling efficiencies, up to {\EEq{\approx}{10}}; higher on average than the ionised phase, although there is limited overlap in this observational sample.

It is important to note that we cannot draw direct comparison between these molecular and ionised phases and our simulated cold/hot outflows, as our simulations do not account for low temperature cooling, molecular chemistry or radiative transfer. 
Instead we focus on qualitative trends for the two gas phases. 
Our simulations predict similar mass outflow rate trends, with our cold phase also dominating at lower {\Lbol} before flattening off at {\LbolEq{>}{46}}. 
Our results for the kinetic coupling efficiency, however, differ from the observational trends. 
We find weak positive (hot phase) and negative (cold phase) trends with {\Lbol}, which are not seen in the observations. Additionally, our values for \EEq{}{} are much lower than those observed. As we have shown, our wind couples weakly to the inhomogeneous ISM and we lose energy to cooling and mixing (Figure \ref{fig:cumulative}). The effect of this can be seen by noting that the cold phase in the smooth simulations has an {\EEq{}{}} an order of magnitude higher than the clumpy case. It is therefore surprising that such high (\EEq{\approx}{1-10}) kinetic coupling efficiencies are found in the observations where we may also expect inhomogeneous ISM conditions.

As we have discussed, there are many observational difficulties and uncertainties when measuring these quantities related to characterising outflow velocities and masses (Section \ref{sec:outflow_v}); radii (Section \ref{sec:radial}); and variations in time (Section \ref{sec:timeevo}) and location (Section \ref{sec:radial_evo}). These challenges could add scatter to the observational results, with additional scatter also being driven by physical differences across individual galaxies such as varying disk masses, ISM distribution and initial wind velocities (see Section \ref{sec:sensitivity}). There are also additional uncertainties that affect outflow properties derived from observations, including constraining electron densities, and more generally, conversion factors between observed line fluxes and total gas masses \citep[e.g.,][]{Rose2018,Lamperti2022,Holden2023a,Holden2023b}, with more recent analysis suggesting that previous work overestimated the outflow rates in the ionised phase by a factor of a few or more, especially in the {\LbolEq{\lesssim}{45}} regime \citep{Davies2020}. 
Additionally, if the observed outflow is assumed to be a spherical shell, it could lead the bulk outflow velocity being overestimated, which could have a large impact on the inferred kinetic luminosity. 

It is also possible that our simulations genuinely under-predict the outflow rate; for example, by neglecting additional driving mechanisms for the outflow, such as radiation pressure, cosmic rays or star-formation. 
Another possibility is that the small-scale wind should be even faster and more energetic than considered here. More massive and centrally-concentrated gas reservoirs or gas configurations with larger covering fractions than considered in our study may also result in more mass-loaded outflows.
Missing physical ingredients, such as metal-line cooling, could increase the cold gas in the outflow. 

In this section, we demonstrated that our simulations predict scaling relations between the mass outflow rate and AGN luminosity. We showed that the scatter on these relations can be affected by the time, disc clumpiness and the minimum radial velocity sensitivity. We compared our results to an observational sample, finding some similarities (such as a turnover in the cold gas outflow rate at {\LbolEq{\gtrsim}{46}}), but also that our simulations have significantly lower kinetic luminosities than those implied from the comparison observational sample.

\section{Conclusions \& Outlook} \label{sec:outlook}

We performed controlled experiments simulating a physically-motivated AGN wind embedded in a clumpy ISM disc. By manually setting the initial ISM structure, we investigated the effect this has on the energetics and multiphase structure of the resulting outflow. We used the \textsc{Arepo} code \citep{Springel2010, Pakmor2016} and the AGN wind model \textsc{Bola} \citep{Costa2020} with AGN luminosities ({\LAGNEq{=}{43-47}}). We divided our results into two main sections. In Section \ref{sec:formation} we investigated the effect of an AGN wind on our clumpy setup and characterised the energetics of the resulting outflow. Our main findings are:

\begin{itemize}
    \item \textbf{Multiscale structure:} the small-scale ultra-fast outflow (UFO) launches large outflow bubbles into the halo, reaching $R=6 \rm{\ kpc}$ by {\tEq{=}{5}} (Figure \ref{fig:summary}). The initial disc inhomogeneities allow the hot gas to vent through high-velocity `chimneys'. The cold outflow is formed from small ($10-20 \rm{\ pc}$), dense ($n \approx 10^{3} \rm{\ cm^{-3}}$) clouds. The venting of the hot gas creates a strong velocity differential between the two phases, with the hot gas streaming at up to {\vrEq{\approx}{1000}}, but the bulk of the cold gas moving at {\vrEq{\approx}{100}} (Figure \ref{fig:vr_hist}). Despite the strong ram pressure this creates on the cold clouds, they are able to survive over the $\tEq{=}{5}$ timescale of the simulations, possibly due to efficient cooling at the phase boundary (Figure \ref{fig:summary}).
    \item \textbf{Multiphase gas energetics:} the outflow is clearly separated into two phases (Figure \ref{fig:phase}), which we define as `hot' (\TEq{>}{4.5}) and `cold' (\TEq{<}{4.5}). The cold outflow carries most of the mass with our fiducial case having {$\dot{M}_{\rm{OF}}=185 \rm{\ M_{\odot}/yr}$} at {\tEq{=}{1}} compared to the hot phase at {$\dot{M}_{\rm{OF}}=40 \rm{\ M_{\odot}/yr}$} (both for {\vminEq{=}{10}}). However, because the hot phase has a much higher velocity, the momentum rates are split roughly evenly between the two phases. In clumpy media, the energy budget is dominated by the hot gas, with around an order of magnitude greater kinetic luminosity (Figure \ref{fig:cumulative}). 
    \item \textbf{Comparison to smooth disc:} we also investigated our AGN wind in a homogeneous disc (Figure \ref{fig:smoothdisc}). We found that the resulting outflow differs significantly from the clumpy case, showing a much narrower spread in radial velocity, with a characteristic speed of {\vrEq{\approx}{400}} in both phases (Figure \ref{fig:vr_hist}). This results in the cold phase of the gas being much more energetic than in the clumpy setup, containing $\sim 70\%$ of the kinetic luminosity of the system at {\tEq{=}{1}} and a higher momentum flux of \pEq{\approx}{7} (Figure {\ref{fig:cumulative}}). 
    \item \textbf{Sensitivity to setup:} we tested the impact on varying the initial galaxy, for example, by reducing the initial density of the disc or altering the sizes of the initial clumps (Section \ref{sec:sensitivity}). We found that the initial clump size made a modest impact, varying the mass outflow rates by up to around $30\%$. Changing the density and height of the disc reduced the mass outflow rates by around a factor of two.
\end{itemize}

In Section \ref{sec:implications} we discussed the implications of our findings for observational studies of AGN outflows. In particular, we found:

\begin{itemize}
    \item \textbf{Outflow measurements:} we discussed the difficulty of observationally measuring outflow properties such as the radius, mass and velocity (Section \ref{sec:challenges_measure}). Many observational studies assume a spherical shell-like outflow, but, as we have seen, an outflow originating from a more realistic, clumpy environment has very different morphology and energetics to a spherical shell. This could lead to incorrect assumptions about the outflow, for example, that all of the gas is moving at a characteristic velocity (Figure~\ref{fig:vr_hist}). This could result in the derived outflow rates being significantly overestimated, especially in colder gas phases.
    \item \textbf{Sensitivity to minimum outflow velocity:} we found that a major source of uncertainty in calculating outflow properties for the cold phase is the minimum radial velocity cut used to define gas as outflowing {\vmin}, which can lead to a factor of 8 difference in mass outflow rate between {\vminEq{=}{10}} and {\vminEq{=}{100}} (Figure \ref{fig:param}). Many observational methods for calculating the outflow rates make an explicit or implicit assumption for {\vmin}. This is particularly challenging when trying to de-couple outflow from non-outflow kinematics (e.g., galaxy rotation) in colder gas phases. A greater value of {\vmin} will lead to a large proportion of the outflowing mass to be missed, with this effect proportionally much worse for the colder phases.
    \item \textbf{Inferring driving mechanisms and kinetic efficiencies:} we found that the, despite our outflow being energy-conserving overall, values of {\pEq{<}{1}} (i.e., momentum-driven) could still be inferred if only a single phase was measured, the full radial extent of the outflow was not captured (Figure \ref{fig:radiusevo}) or the outflow was observed past the momentum peak (Figure \ref{fig:timeevo}). This makes it difficult for observations to accurately determine the driving forces behind any outflow seen (Figure \ref{fig:mom_vel}). Additionally, the derived kinetic energy coupling efficiencies were seen to be highly dependent on the phase, time and location of the outflow (Figures \ref{fig:radiusevo} and \ref{fig:timeevo}). This makes inferences about the efficiency of the wind from large-scale measurements of the kinetic coupling efficiency ({\Ek}) challenging. 
    \item  \textbf{Scaling relations:} we found that our simulations predict a positive correlation between the mass outflow rate and AGN luminosity in both the hot and cold phase (Figure \ref{fig:Fiore_ME}). The cold phase dominates at lower {\Lbol}, but flattens off at {\LbolEq{>}{46}}. This turnover point is also seen in the observational compilation of \cite{Bischetti2019}. However, we find lower kinetic coupling efficiencies than observed, especially in the cold phase. Future work warrants a more comprehensive comparison to observations, including accounting for systematic uncertainties and other potential sources of scatter (see Section \ref{sec:fiore}). 
\end{itemize}

The observation of cold clouds entrained in galactic outflows \citep[e.g.,][]{DiTeodoro2019,Veilleux2020} is puzzling as the typical cloud crushing timescale is smaller than the outflow timescale \citep{Klein1994,Zhang2017,Schneider2017}. In our simulations, we do find cold gas clouds surviving on Myr timescales. Cold, dense gas can be fast, travelling at velocities up to $800 \, \rm km \, s^{-1}$ (see Figure~\ref{fig:phase}), but does not appear to reach the extreme velocities $> 1000 \, \rm km \, s^{-1}$ suggested by some observations \citep[e.g.][]{Lutz2020}.
A full study of the formation and evolution of the cold clouds in our simulations is beyond the scope of this paper, but constitutes an important future work direction.

In this study we only considered primordial cooling down to {\TEq{\approx}{4}}. Metal-line cooling boosts the cooling rate at \TEq{\approx}{5-7} and allows cooling to {\TEq{<}{4}}. Our values for mass outflow rates in the cold component can therefore be considered conservative lower estimates}. In particular, how much of our `cold phase' eventually turns molecular remains to be understood. Our estimated outflows rates $\approx 10^{2-3} \rm{\ M_\odot}$ still fall short of observational reports of outflow rates $> 10^3 \, \rm M_\odot$ \citep{Fiore2017} even for our brightest simulated AGN. Performing new simulations with metal-line and low temperature cooling is will thus be important in future studies. Furthermore, our initial cold clumps have densities up to $n \lesssim 10^{3}\ \rm{cm^{-3}}$ which may not capture the highest densities seen in the cores of molecular clouds ($n\approx 10^{2-6}\ \rm{cm^{-3}}$; \citealt{Ferriere2001}). We showed in Figure \ref{fig:param} that reducing the mean initial density resulted in less cold outflowing gas, but slightly more hot gas. Raising the density could therefore increase the cold phase outflow rate, which could result in a closer match to the observations (Figure \ref{fig:Fiore_ME}), however, increasing the density of the disc is unlikely to result in a higher outflow velocity for the cold gas, which is the main cause of the differences we see with the observations.

There are additional physical processes neglected here that will have to be examined in future studies. These include magnetic fields, AGN radiation and cosmic rays. Previous work on single-cloud simulations including magneto-hydrodynamics, for instance, has found that magnetic fields can have a range of effects, for example, enhancing thermal instabilities \citep{Ji2018}, or perhaps suppressing them \citep{Gronke2020a}. In a recent study, \cite{Hidalgo-Pineda2024} found that the interplay between magnetic fields and radiative cooling can reduce the size of entrained cold clouds and lead to more rapid entrainment. In the case of cosmic rays, the pressured built around cold clouds via the `bottleneck effect' may play a major role in producing cold gas outflows \citep[e.g.][]{Bruggen:20}. 

Since we find radiative cooling operates both in shocks and in mixing layers (Figure \ref{fig:summary}), it will be important to further understand the observational imprints of the associated cooling emission. Since mixed gas has a temperature of \TEq{\approx}{6} in our simulations, it is possible this produces extended X-ray emission. Interestingly, there are X-ray `chimneys' in our own Milky Way linking the galactic nucleus to the Fermi bubbles \citep{Ponti2019,Ponti2021} resembling the low-density gaps through which hot gas vents through in our simulations. Generating multi-wavelength predictions based on our simulations will be important to test AGN feedback through winds.

\section*{Acknowledgements}

We thank the referee for a helpful and constructive report, which improved the clarity of the paper. S.R.W. acknowledges funding from the Excellence Cluster ORIGINS and computing facilities from the Computational Center for Particle and Astrophysics (C2PAP). The ORIGINS cluster is funded by the Deutsche Forschungsgemeinschaft (DFG; German Research Foundation) under Germany's Excellence Strategy: EXC-2094-390783311. C.M.H. acknowledges funding from a United Kingdom Research and Innovation grant (code: MR/V022830/1). We thank Manuela Bischetti for sharing their observational data. This work also used the DiRAC Memory Intensive service (Cosma8) at Durham University, managed by the Institute for Computational Cosmology on behalf of the STFC DiRAC HPC Facility (www.dirac.ac.uk). The DiRAC service at Durham was funded by BEIS, UKRI and STFC capital funding, Durham University and STFC operations grants. DiRAC is part of the UKRI Digital Research Infrastructure.

\section*{Data Availability}
The data underlying this paper will be shared on reasonable request
to the corresponding author.


\bibliographystyle{mnras}
\bibliography{lib.bib}


\appendix

\section{Numerical Convergence} \label{ap_resolution}

\begin{figure}
    \centering
    \includegraphics[width=0.35\textwidth]{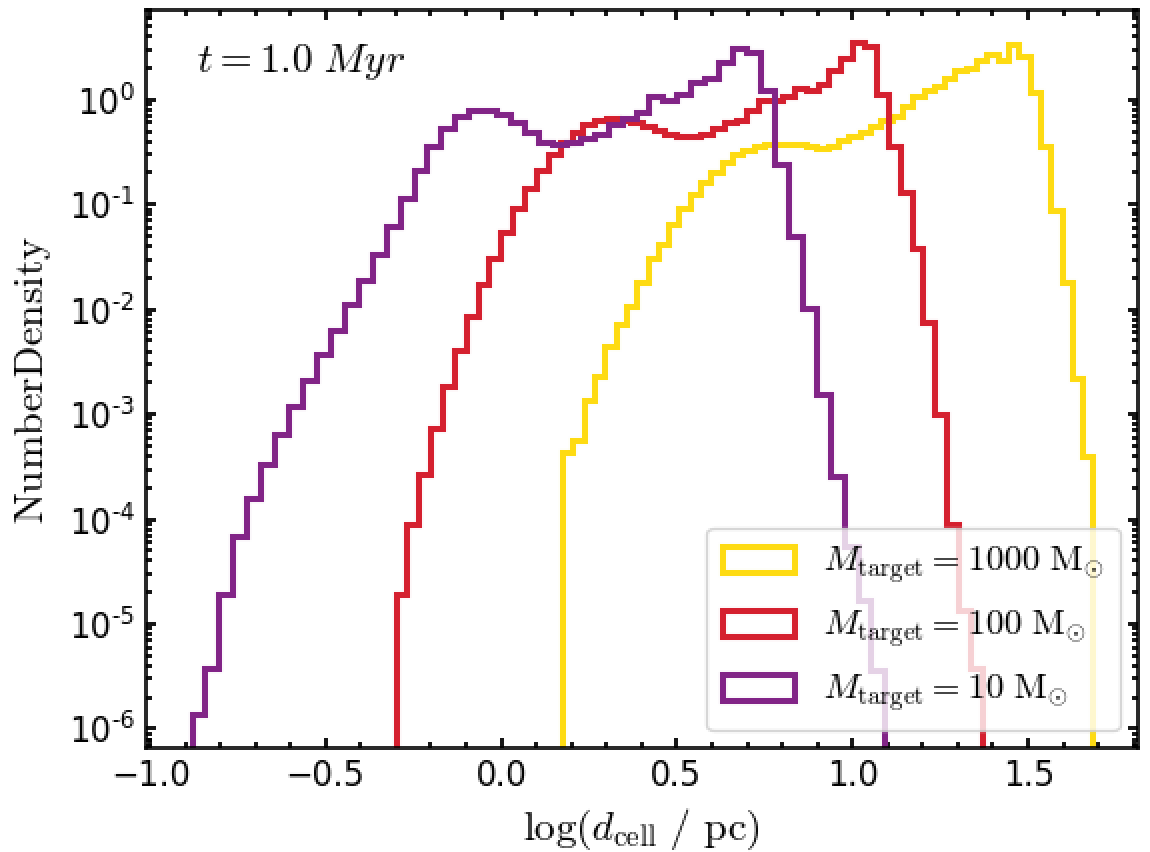}
        \caption{The distribution of cell diameters in our three resolution test simulations at \tEq{=}{1}. The minimum cell diameter for each resolution is roughly $d_{\rm{cell, min}}\approx 1.5, 0.5, 0.2 \rm{\ pc}$, for our target mass resolutions of \MtargetEq{=}{1000,100,10} respectively.}
    \label{fig:app_cell}
\end{figure}

To test the numerical convergence of our results, we performed simulations at target mass resolutions of {\MtargetEq{=}{1000}} (`low resolution'), {\MtargetEq{=}{100}} (`fiducial resolution') and {\MtargetEq{=}{10}} (`high resolution'). 
These were performed with our fiducial parameters, i.e., a disc clumpiness of {\LmaxEq{=}{170}} ({\kminEq{=}{12}}) and AGN luminosity of {\LAGNEq{=}{45}} (see Table \ref{table:runs} for a full description of our fiducial model). Due to the significant computational cost of the high resolution simulation, we only perform it to \tEq{=}{1}. 
Figure \ref{fig:app_cell} shows the distribution of the cell diameters for the three resolution test simulations, assuming a spherical geometry for the Voronoi cells. Improving the mass resolution by a factor of 10 should improve the spatial resolution by a factor of $\sqrt[3]{10} \approx 2.2$ which is broadly in line with what we see here. Of particular interest is the minimum cell diameter probed as this will be in the densest gas where we probe the structure of the entrained cold gas clouds. The low/fiducial/high resolution simulations have minimum spatial resolutions of $d_{\rm{cell, min}}\approx 1.5, 0.5, 0.2 \rm{\ pc}$ respectively. Our fiducial simulation has higher resolution than other similar simulations of AGN interacting in idealised clumpy discs (e.g., \citealt{Mukherjee2018} and \citealt{Tanner2022} use grid-based systems with resolutions of 6 pc and 10 pc respectively) and our high-resolution simulation offers unprecedented resolution for galaxy-scale simulations.

\begin{figure}
    \centering
    \includegraphics[width=0.35\textwidth]{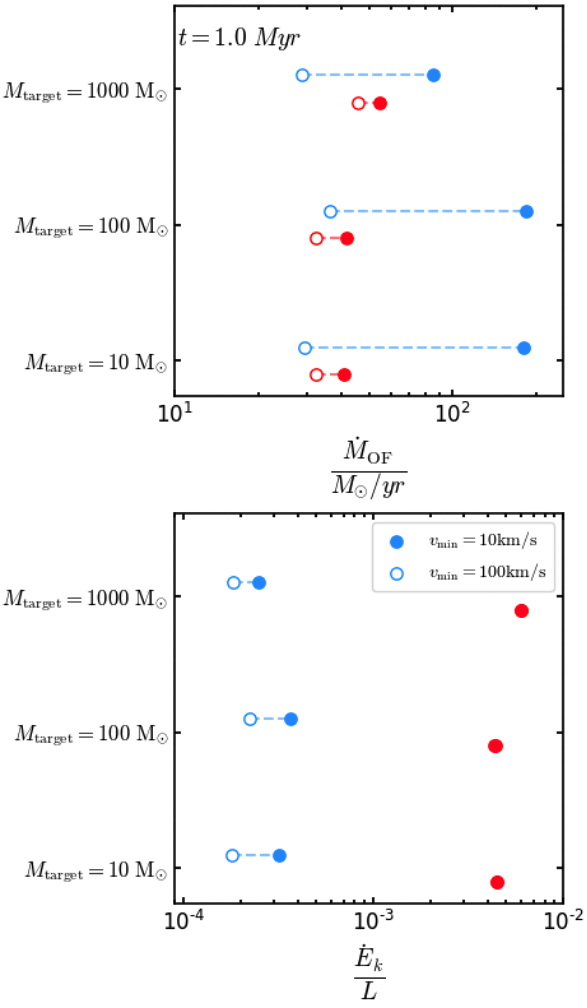}
        \caption{Convergence of global properties, adapted from Figure \ref{fig:param}. We show the mass outflow rate (top) and kinetic luminosity (bottom) for the three resolutions (\MtargetEq{=}{10,100,1000}). The blue/red points represent our cold/hot phases, and the solid/hollow points show minimum radial velocity cuts of \vminEq{=}{10} and \vminEq{=}{100}, respectively. We can see that there is little change in the global outflow properties between the fiducial and high resolution simulations, demonstrating these properties have converged at \MtargetEq{=}{100}.}
    \label{fig:app_global}
\end{figure}

\begin{figure*}
    \centering
    \includegraphics[width=0.8\textwidth]{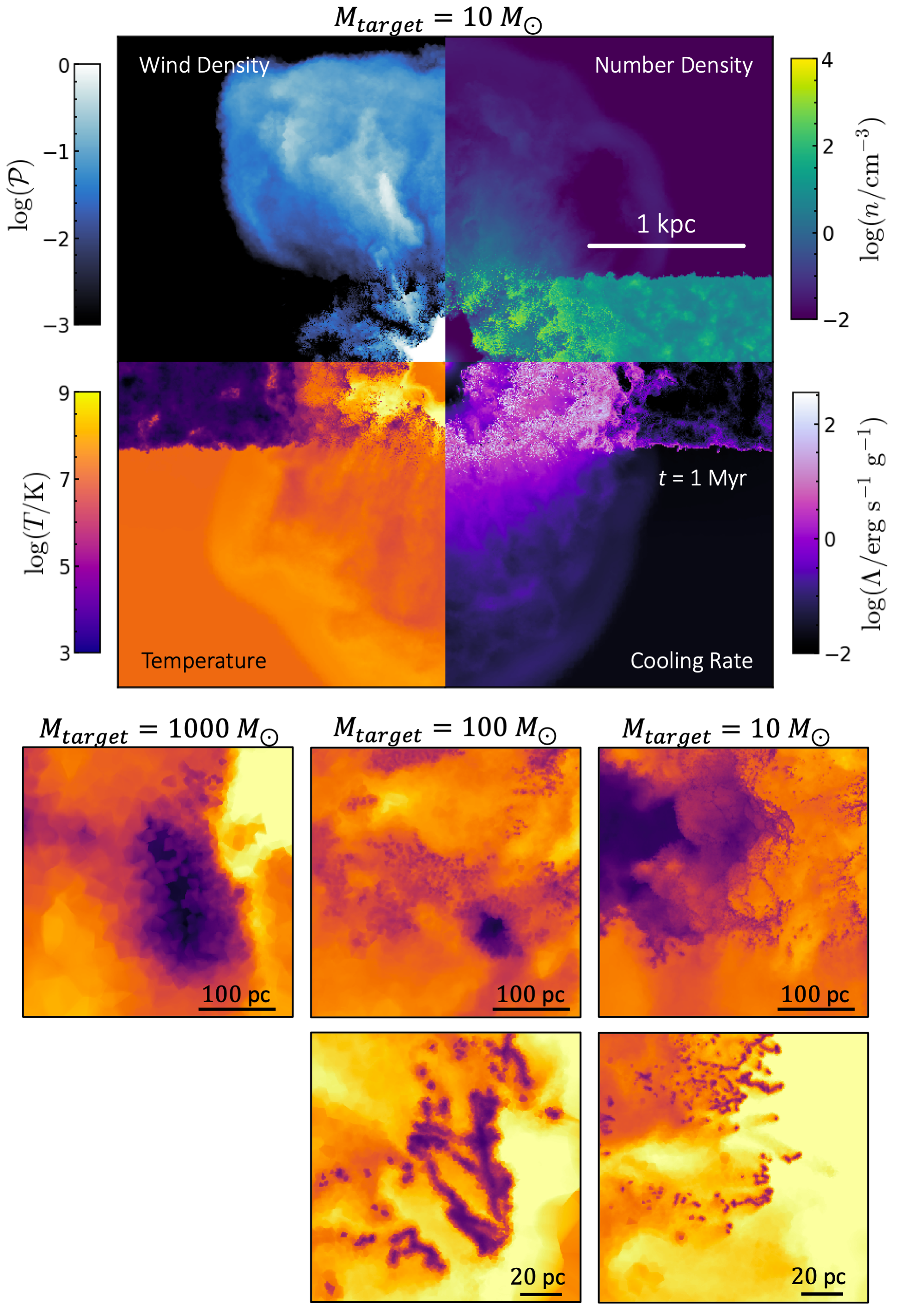}
        \caption{The results of our high-resolution simulation (\MtargetEq{=}{10}). The top panel shows the galaxy edge-on at \tEq{=}{1}. We show (clockwise from top left) the wind tracer density, gas number density, cooling rate and temperature. The outflow at this spatial scale is broadly similar to that of our fiducial resolution, shown in Figure \ref{fig:summary}. At the bottom of the Figure, we show a comparison of the structure of the small-scale cold gas clouds for the three resolution simulations. The columns show the resolution (\MtargetEq{=}{1000,100,10}) and the rows show two different spatial scales, shown by the scale bars. The fiducial resolution simulation shows cold clouds of sizes down to $\approx 10 \rm{\ pc}$, but in the high resolution simulation has even smaller clouds, down to $\approx 1 \rm{\ pc}$. This emphasises the difficulty in resolving the cold gas phase in galaxy-scale simulations.}
    \label{fig:app_clumps}
\end{figure*}

To investigate the numerical convergence of our global outflow properties, we computed the mass outflow rate and kinetic energy coupling efficiency for our three resolution tests, shown in Figure \ref{fig:app_global}. The symbols are the same as Figure \ref{fig:param}, with blue/red showing the cold/hot phase and the filled/hollow markers showing the effect of varying the minimum outflow velocity (\vmin). We can see that the low resolution simulations (top) has lower mass outflow rates and kinetic energies in the cold phase than the fiducial simulation (middle), but slightly higher values in the hot phase. This could be because the larger cold gas clumps seen in the low resolution simulation (Figure \ref{fig:app_clumps}) are harder to accelerate and have a lower surface area resulting in less phase mixing and cooling than in the fiducial simulation. However, the differences between the fiducial simulation and the high resolution simulation (bottom) are much smaller: the hot phase is essentially the same, and the cold phase only differs by a slight decrease in mass outflow rate for {\vminEq{=}{100}}. This demonstrates that the global outflow properties are well-converged at our fiducial resolution.

In Figure \ref{fig:app_clumps} we show a similar plot to Figure \ref{fig:summary} for our high-resolution simulation at {\tEq{=}{1}}. The top section shows an edge-on view of our disc showing (clockwise from top left) the wind density, number density, cooling rate and temperature. At this kpc-scale, the high-resolution outflow looks broadly similar to the fiducial resolution seen in Figure \ref{fig:summary}. This corroborates our finding in Figure \ref{fig:app_global} that the global outflow is well-converged at our fiducial resolution. However, the bottom section of this panel shows the morphology of the entrained cold clouds the three different resolutions (columns) at different spatial scales (rows). We can see that increasing the resolution decreases the size of the smallest cloudlets seen: at our fiducial resolution (\MtargetEq{=}{100}; middle column), the smallest clouds are on the scale of $\approx 10 \rm{\ pc}$ whereas in the high resolution simulation (\MtargetEq{=}{10}; right column), they can be as small as $\approx 1 \rm{\ pc}$. The shapes of the cold gas clouds are similar, showing initially dense cores surrounded by smaller fragments and filaments, but the scale clearly varies with resolution. However, even high-resolution simulations of single clouds interacting with a wind do not always find convergence, with \cite{Yirak2010} finding that a ratio of cloud-to-resolution element ratio of $r_{\rm{cl}}/d_{\rm{cell}}>100$ is not always sufficient for self-convergence. Conversely, \cite{Gronke2020a} and \cite{BandaBarragan2021} argue that their global quantities (e.g., mass entrainment rate) converge at $r_{\rm{cl}}/d_{\rm{cell}} \geq 8$, even if the exact morphology of the cold fragments requires much higher resolution to converge. This feature of global properties converging while the structure of the gas continues to granulate may be a common and unavoidable feature of the cold gas phase \citep{Hummels2019,vandeVoort2019,Nelson2020}. Nevertheless, as we have shown, the exact spatial structure of the cold clouds does not affect the global outflow properties, which is the focus of this work.

\bsp	
\label{lastpage}
\end{document}